\documentclass[preprint,12pt, 3p , times]{elsarticle}

\usepackage[]{hyperref}

\usepackage{amssymb}
\usepackage{amsmath}

\usepackage{xcolor}
\usepackage{lineno}

\usepackage{graphicx}       
\usepackage{listings}

\definecolor{codegreen}{rgb}{0,0.6,0}
\definecolor{codegray}{rgb}{0.5,0.5,0.5}
\definecolor{codepurple}{rgb}{0.58,0,0.82}
\definecolor{backcolour}{rgb}{0.95,0.95,0.92}
\definecolor{LightGray}{gray}{0.95}
\definecolor{darkgrey}{rgb}{0.2, 0.2, 0.2}
\lstMakeShortInline[columns=fixed]|

\journal{Computer Physics Communications}

\newcommand*{\hepaid}{{\color{teal}\texttt{he\-p-a\-id}}}

\definecolor{bg}{rgb}{0.1,0.1,0.1}  
\definecolor{fg}{rgb}{1,1,1}        
\definecolor{keyword}{rgb}{0.8,0.2,0.2} 
\definecolor{comment}{rgb}{0.5,0.5,0.5} 
\definecolor{string}{rgb}{0.2,0.6,0.2}  
\begin{document}

\begin{frontmatter}



\title{\hepaid: A Python Library for Sample Efficient Parameter Scans in Beyond the Standard Model Phenomenology}

\author[label1]{Mauricio A. Diaz} 
\author[label2]{Srinandan Dasmahapatra}
\author[label1,label3]{Stefano Moretti}

\affiliation[label1]{organization={School of Physics \& Astronomy, University of Southampton},
            city={Southampton},
            postcode={ SO17 1BJ},
            country={UK}}

\affiliation[label2]{organization={School of Electronics \& Computer Science, University of Southampton},
            city={Southampton},
            postcode={SO17 1BJ},
            country={UK}}
            
\affiliation[label3]{organization={Department of Physics \& Astronomy, Uppsala University},
            addressline={Box 516}, 
            postcode={75120},
            city={Uppsala},
            country={Sweden}}


\begin{abstract}
This paper presents \hepaid, a modular Python library conceived for utilising, implementing, and
developing parameter scan algorithms. Originally devised for sample-efficient, multi-objective
active search approaches in computationally expensive Beyond Standard Model (BSM) phenomenology,
the library currently integrates three Machine Learning (ML)-based approaches: a Constraint Active
Search (CAS) algorithm, a multi-objective Active Search (AS) method (called b-CASTOR), and a self-exploration method named Machine Learning Scan (MLScan). These approaches address the challenge of
multi-objective optimisation in high-dimensional BSM scenarios by employing surrogate models and
strategically exploring parameter spaces to identify regions that satisfy complex objectives with
fewer evaluations. Additionally, a Markov-Chain Monte Carlo method using the Metropolis-Hastings algorithm (MCMC-MH) is implemented for method comparison. The library also
includes a High Energy Physics (HEP) module based on SPheno as the spectrum calculator. However,
the library modules and functionalities are designed to be easily extended and used also with other  
external software for phenomenology. This manual provides an introduction on how to use the
main functionalities of \hepaid\ and describes the design and structure of the library.
Demonstrations based on the aforementioned parameter scan methods  show that \hepaid\ methodologies enhance the
efficiency of BSM studies, offering a versatile toolset for complex, multi-objective searches for new physics  in
HEP contexts exploiting advanced ML-based approaches.

\end{abstract}




\end{frontmatter}


\tableofcontents

\section{Introduction}
Within Beyond Standard Model (BSM) phenomenology, the Parameter Scan (PS) problem \cite{Brein_2005} involves a systematic
exploration of the multi-dimensional parameter space of a new physics scenario. This process includes calculating
numerical values for model predictions across various points in its parameter space, applying
experimental and theoretical constraints, and identifying satisfactory regions that can explain
multiple phenomena. The satisfactory regions in the parameter space are found by checking whether
a theoretical prediction matches within some error margin measured  features of anomalous data or
respects exclusion limits if no BSM observations have been made.

PS methods must address several computational challenges, covering
high-dimensional parameter spaces, the computational cost of numerical evaluations, and satisfying a large
number of constraints in the physical observable space. Additionally,  PS methods may be
further limited by the computational resources available for the study. Therefore, selecting a PS
method suitable for a specific phenomenological study is not trivial. Expert knowledge of the BSM
model and its computational demands must be evaluated to perform a successful phenomenological study.

The PS problem is commonly framed as a sampling problem \cite{Hogg_2018},
tackled using Bayesian inference techniques \cite{albert2024comparison}, such as 
Markov-Chain Monte Carlo (MCMC) methods (see, e.g., \cite{Speagle:2019ffr}), to estimate the probability density functions across the parameter
space. Recent advances have explored the integration of Machine Learning (ML)
\cite{feickert2021living,Baruah_2024} methods into PS techniques offering a promising
approach to addressing challenges related to efficiency and scalability. Neural Network (NN)-based methods \cite{ren2019exploring, Hammad_2023, binjonaid2024multilabel,Hollingsworth_2021}
have been investigated by framing the problem as  a regression, classification, or
generative task. This involves training a NN model to approximate the observables from the
inputs or to classify and assess the viability of parameter configurations under experimental
constraints. These models are subsequently incorporated into informed sampling strategies. Active
Learning (AL) \cite{Goodsell_2023, Goodsell2023BSMArtSA} techniques have also been utilised to
train NNs to approximate the decision boundaries for the allowed parameter space. Furthermore,
the PS problem has been framed for black-box optimisation techniques, employing
methods such as Bayesian optimisation and evolutionary strategies \cite{PhysRevD.107.035004,
romao2024combining, balazs_comparison_2021}. Recently, another formulation was investigated by
applying Active Search (AS) approaches to enhance sample efficiency in computationally
expensive BSM scenarios \cite{madiaz2024}. This kind of work led to the development of the library
introduced here.

Several PS and sampling libraries have been developed for phenomenology, including
{\tt BSM Toolbox} \cite{Staub:2011dp}, {\tt xBit} \cite{Staub:2019xhl}, \texttt{EasyScan\_HEP} \cite{Shang:2023gfy}, and
{\tt BSMart} \cite{Goodsell:2023iac}. Each of these libraries addresses the PS  problem with
unique software designs and specific usage goals. They share common features such as integration
with a set of High Energy Physics (HEP) packages, implementation of various PS 
algorithms, and the use of configuration files to simplify setup. These tools provide the
community with a range of resources tailored to different applications.

In this work, we introduce a new Python library, \hepaid, which provides a modular
framework for developing PS algorithms for BSM phenomenology and also adopts the
principles of related libraries. It manages the  HEP software and provides
components to ease the utilisation, implementation, and development of PS algorithms
for phenomenological studies. The library comprises two main modules: the \texttt{hep} module and
the \texttt{search} module.

The \texttt{hep} module facilitates the integration of the HEP software ecosystem into Python
scripts. It allows users to perform a first-principles computation of observable quantities to compare with experimental data for each parameter space point
using a stack of HEP software, collecting the output with a single function call. Currently,
the SARAH \cite{Staub:2008uz,Staub_2014} family of programs is implemented. The \texttt{search}
module manages PS algorithms, following an AS \cite{garnett2012bayesian} paradigm in
which a search policy and a surrogate model are employed to explore the parameter space of a
multi-objective function to find parameter configurations where the objectives satisfy a set
of constraints. This framework allows the integration of potentially any PS method, such as MCMC
or ML based sampling methods. The connection between the PS algorithms in the
\texttt{search} module and the HEP software in the \texttt{hep} module is established through the
construction of an \textit{objective function}. The \texttt{search} module includes an
\textit{objective function} constructor, which defines the search space, objectives, and
constraints based on a predefined configuration. It also maintains an internal dataset of samples
with functionalities for saving, loading, and exporting datasets in formats such as a Pandas DataFrame 
\cite{reback2020pandas}  or a PyTorch \cite{paszke2019pytorch} tensor.

We demonstrate the use of the library through quick tests and real BSM scan examples. We employ
an AS parameter scan method to study the $(B-L)$ Supersymmetric Standard Model
($(B-L)$SSM) \cite{Abdelalim:2020xfk}, accommodating results from an
observed signal at approximately 95 GeV in neutral (pseudo)scalar searches in the \( h \rightarrow \gamma
\gamma \) channel, as observed by CMS \cite{CMS:2023yay}.

The article is structured as follows. In Section \ref{sec:overview}, we provide a brief overview
of the library, followed by code examples for different use cases. (This section also serves as a quick
start guide.) In Section \ref{sec:hepmodule}, we describe the integration of HEP software, i.e., the \texttt{hep} component. Section
\ref{sec:methods} explains the internal structure of the \texttt{search} component of the
library and details the parameter scans implemented in \hepaid. We then conclude
in Section \ref{sec:summa}.

\section{The \texttt{hep-aid} Library}\label{sec:overview}

\begin{figure}[h]
  \centering
  \includegraphics[width=0.7\textwidth]{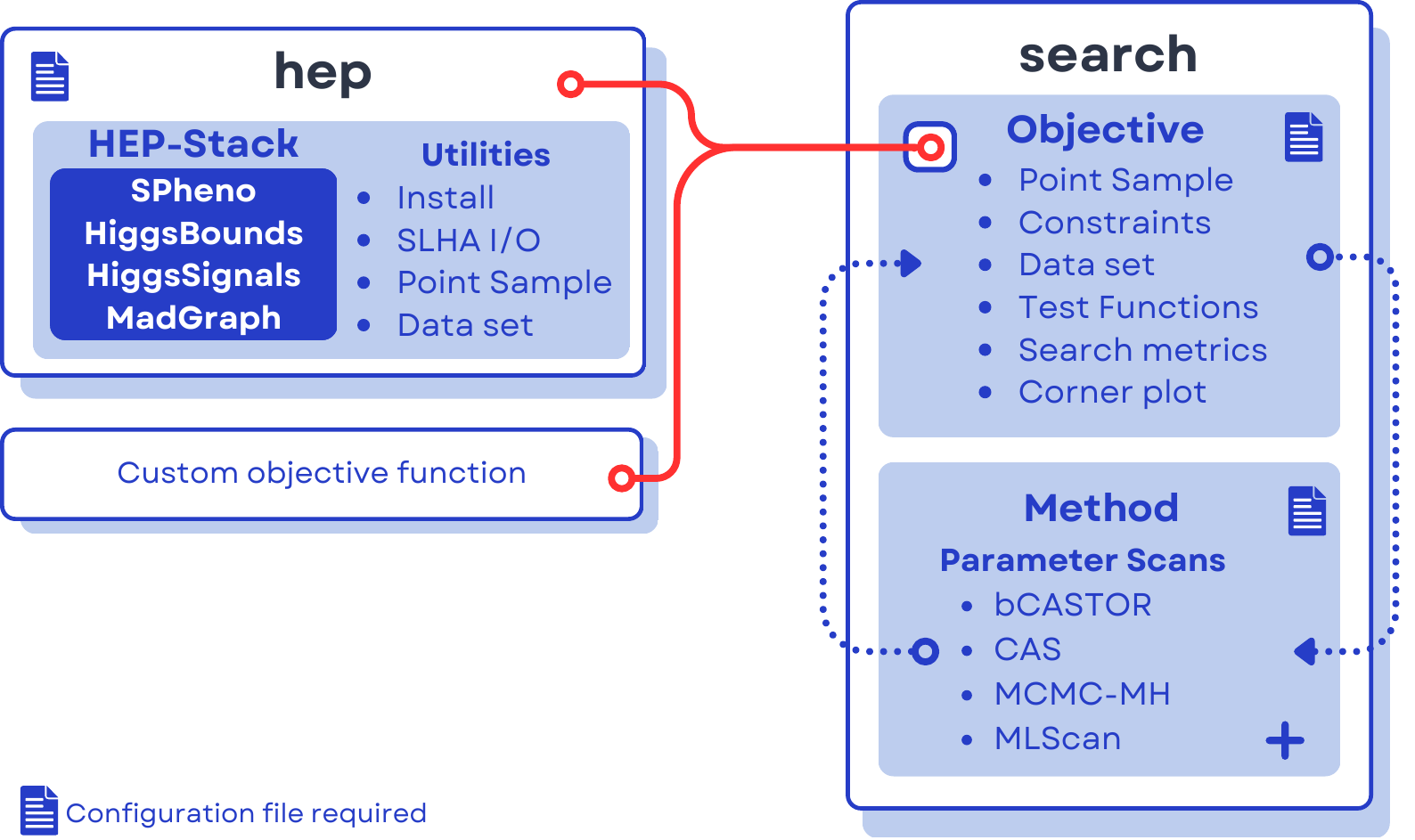}  
  \caption{Diagram showing the general structure of the \hepaid\ library, which includes two main
    modules: {\tt hep} and {\tt search}. The search module contains the objective and methods submodules. The
    {\tt objective} submodule allows users to plug in any custom objective function. The {\tt hep} module
    serves as a specialised external objective function that integrates with HEP 
    software and can be seamlessly used by the objective submodule for integration.}
  \label{fig:hepaid}
\end{figure}

The \hepaid\ library  provides a modular framework for performing parameter scans in BSM scenarios,
currently using SPheno \cite{Porod2003SPhenoAP,Porod_2012}, HiggsBounds (HB) \cite{Bechtle_2010},
HiggsSignals (HS) \cite{Bechtle_2014} and MadGraph (MG) \cite{Alwall:2014hca}, which we call the \textit{HEP-stack}.
It is focused on the AS \cite{garnett2012bayesian} paradigm, where a multi-dimensional multi-objective function
needs to be defined and the PS algorithm searches for
satisfactory configurations in the parameter space given a set of constraints on the objectives.
The search is done using a surrogate model, which is fitted to the collected data, to
approximate the objective function and to assess which regions of parameter space to explore by querying the {HEP-stack}. This yields parameter
configurations where the objectives satisfy the constraints. We call this region the
\textit{satisfactory region}. Originally the library was created to give the user simple access to use the
b-CASTOR \cite{madiaz2024} and Constraint AS (CAS) \cite{malkomes2021beyond} algorithms for parameter scans but,
given the modular structure of it, many parameter scan algorithms can be implemented: e.g., 
 \hepaid\ already includes a MCMC method using the Metropolis-Hastings sampling algorithm (MCMC-MH) \cite{Metropolis:1953am,Hastings:1970aa}, MLScan
\cite{ren2019exploring}, and a NN based sampling algorithm for BSM phenomenology. The
library is divided into two key modules, the \texttt{hep} module and the \texttt{search} module,
illustrated in Figure \ref{fig:hepaid}.

The \texttt{hep} module facilitates the integration of the HEP software ecosystem into Python
scripts. The \texttt{hep.stack}  module integrates SPheno, HB, HS and MG
in a sequential stack of software. Technically, this has four {HEP-stacks}, ranging from SPheno used independently to the full chain incorporating all four HEP programs, and the user can utilise any of these {HEP-stacks} depending on the
phenomenological study. Each {HEP-stack} includes operational utilities, assisting in handling and
managing data, mainly SUSY Les Houches Accord (SLHA) \cite{Allanach:2008qq} files, and running the software externally. The HEP-stacks need to be
initialised with a pre-defined configuration file. This configuration file contains information
about the SPheno inputs that will undergo a parameter scan and the necessary
directories for the HEP tools used in the scan.

The library offers a quick installation method integrated as a command-line feature, along with a
function to create a template HEP-stack configuration file. Once the configuration file is
defined and the HEP-stack is initiated, the \texttt{hep.stack} module enables the user to run a
parameter space configuration defined as a simple array. This array corresponds to the input
parameters for SPheno, as specified in the configuration file. 
The result contains all the input and output
information of the HEP tools used in the HEP-stack, formatted as a Python dictionary.

\begin{figure}
    \centering
    \includegraphics[width=0.6\linewidth]{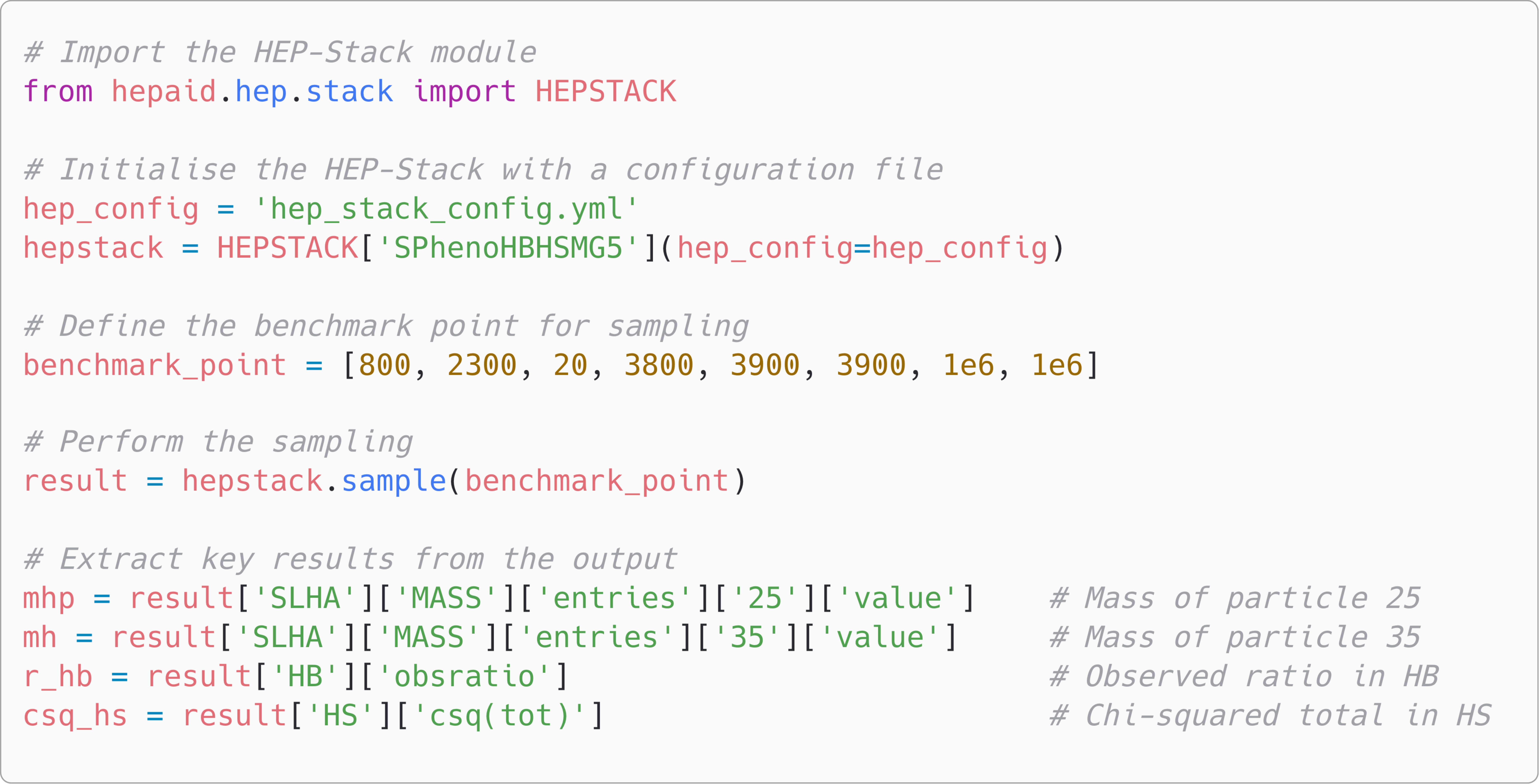}
    \caption{Example code demonstrating the initialisation of the HEP-stack with SPheno,
    HB, HS, and MG using a HEP-stack class specific to the
    ($B-L$)SSM. The corresponding \texttt{SPhenoHBHSMG5} object runs a parameter configuration defined by
    an array, returning results in a Python dictionary. Key outputs, such as the Higgs particle
    masses (\texttt{mhp}, \texttt{mh}), the HB ratio (\texttt{r\_hb}), and the HS
    $\chi^2$ (\texttt{csq\_hs}), can be extracted for analysis.}
    \label{fig:using_hep_stack}
\end{figure}

The \texttt{search} module provides all the necessary components for performing a search on a multi-objective function, either by using the \texttt{hep} module or by defining a custom function. It also includes PS algorithms and tools to support experimentation and the development of new PS methods. Since we adhere to the
AS approach, \hepaid\ also includes \textit{surrogate models} to approximate the objective
function under study. These features are implemented in three main sub modules,
\texttt{search.objective}, \texttt{search.methods} and \texttt{search.models}. The \texttt{objective}
module contains the \texttt{Objective} class, which is the objective function constructor it
needs to be initialised with a configuration file stating the search space dimensions with their
ranges, the objectives variables and the information of the constraints. The \texttt{Objective}
class internally manages the sampling of the objective function, stores the dataset, performs
data processing to export to the \textit{surrogate models}, and saves the dataset to disk. The
\texttt{methods} module contains all the PS methods available currently, namely, MCMC-MH \cite{Metropolis:1953am,Hastings:1970aa},
b-CASTOR  \cite{madiaz2024}, CAS \cite{malkomes2021beyond} and MLScan \cite{ren2019exploring}.
Each method is constructed by inheriting the \texttt{Method} base class which loads a specific configuration file and saves or loads checkpoints to continue the search. A set of metrics is recorded in each PS method, such as the total number of parameters sampled, and those that satisfy the constraints. Further metrics may be customised for different use cases. Finally, if the parameter scan algorithms follow an AS approach, they will use
the surrogate models implemented in the \texttt{search.models} module currently containing Gaussian
Processes (GPs) for the AS algorithms and a simple Multi-Layer Perceptron (MLP) for the MLScan method. 

The workflow idea that \hepaid\ proposes is that the users define an objective function using the
\texttt{Objective} class and define its configuration file. They can then run the preferred PS method which uses the initiated {\tt Objective} object. 
For BSM phenomenological analyses using the \texttt{hep} module, the workflow needs to be more elaborated
since the objective function needs to be constructed in a appropriate manner compatible with the
\texttt{Objective} class, by retrieving the necessary values of  masses, cross sections, Branching Ratios (BRs), etc.: see   the code example in Figure \ref{fig:using_hep_stack}.


\subsection{Library Overview}
Here, we describe in detail how to make use of \hepaid.

\subsubsection{Installation}
\begin{figure}[!h]
    \centering
    \vspace{-10mm}
    \includegraphics[width=0.5\linewidth]{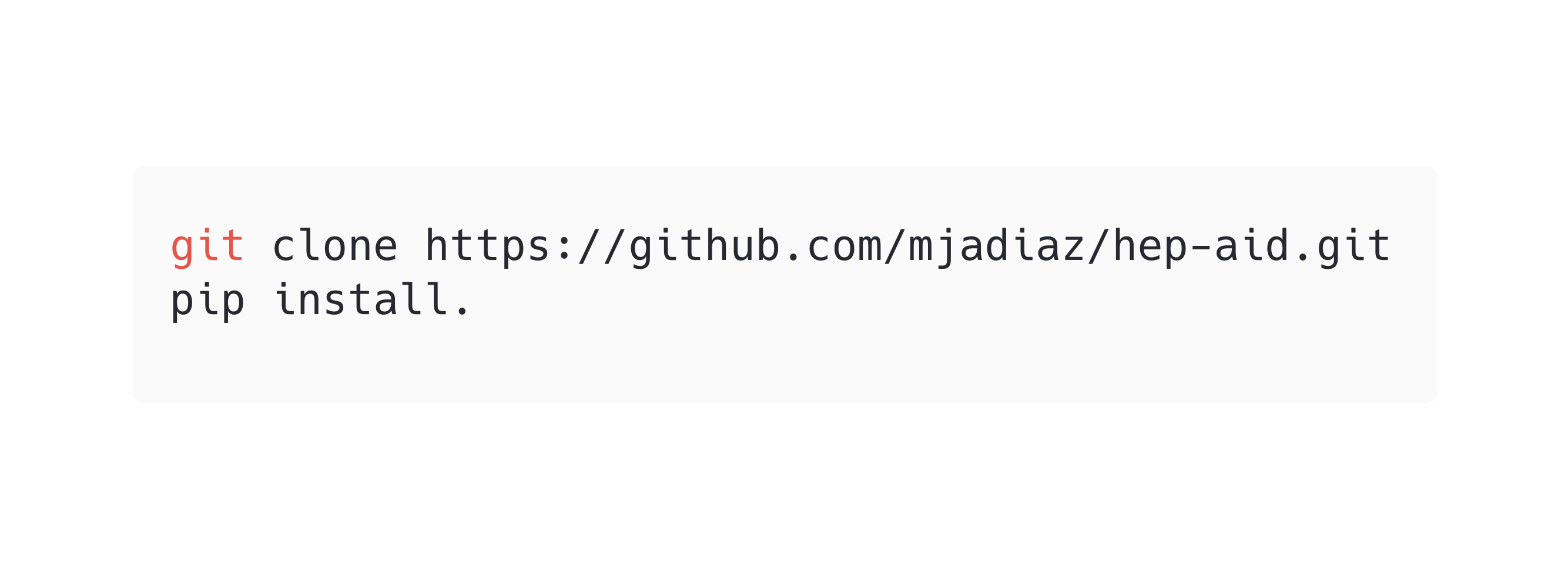}
    \vspace{-12mm}
    \caption{Commands to clone the repository and install the \hepaid\ package.}
    \label{fig:cloneinstall}
\end{figure}

The library is publicly available in GitHub\footnote{https://github.com/mjadiaz/hep-aid.}. To
install \hepaid\ one needs to clone the repository first, then proceed as illustrated in Figure \ref{fig:cloneinstall}.
\subsubsection{Test Objective Functions}
\begin{figure}[!h]
    \centering
    \includegraphics[width=0.7\linewidth]{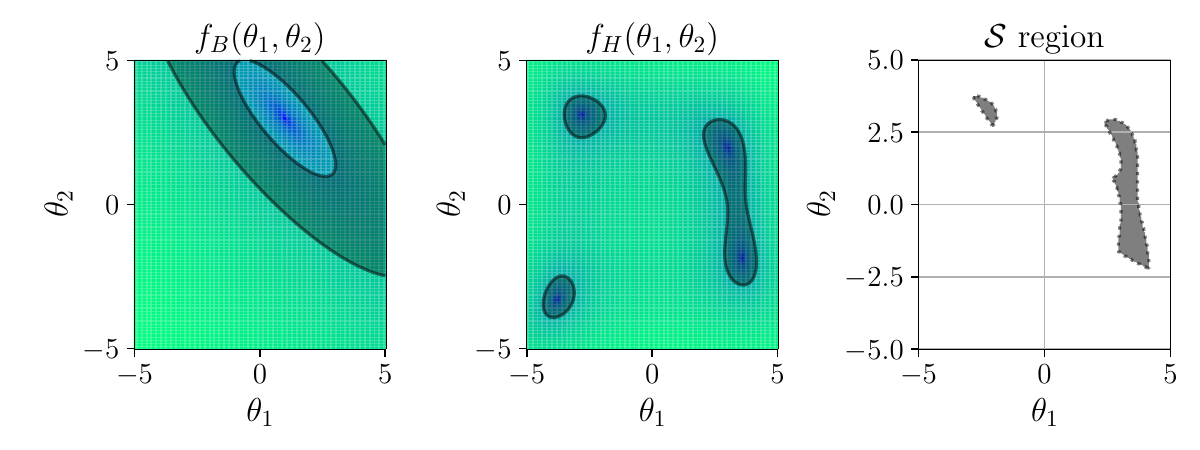}
    \caption{Two-dimensional double-objective test function, where the objectives must satisfy
    the shaded areas. The overlapping shaded areas define the satisfactory region, denoted by
    $\mathcal{S}$ on the right plot.}
    \label{fig:test_truth}
\end{figure}

\begin{figure}[!h]
    \centering
    \includegraphics[width=0.8\linewidth]{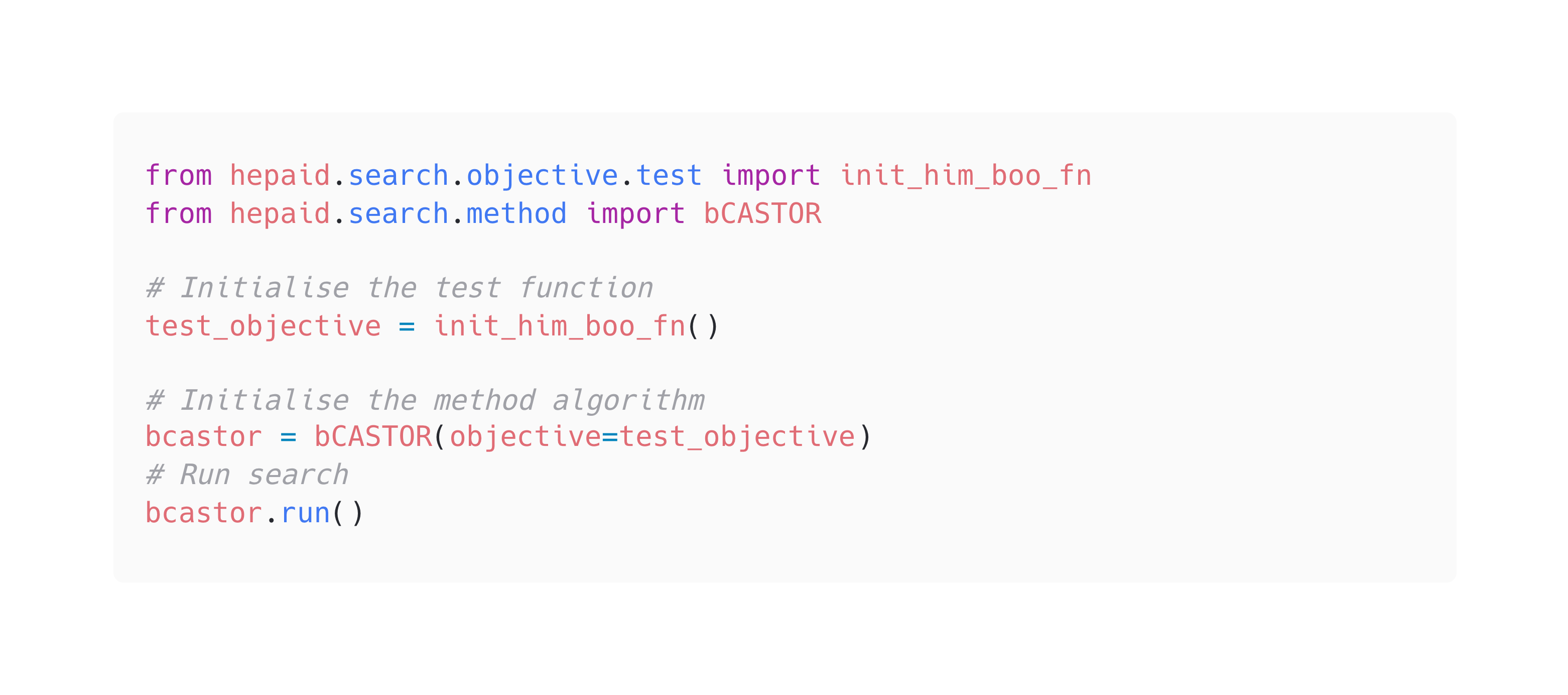}
    \vspace{-12mm}
    \caption{Quick start guide illustrating the following: initialising a two-dimensional double objective test function, selecting and starting the b-CASTOR method, then running the search.}
    \label{fig:quick1}
\end{figure}

\begin{figure}
    \centering
    \includegraphics[width=0.8\textwidth]{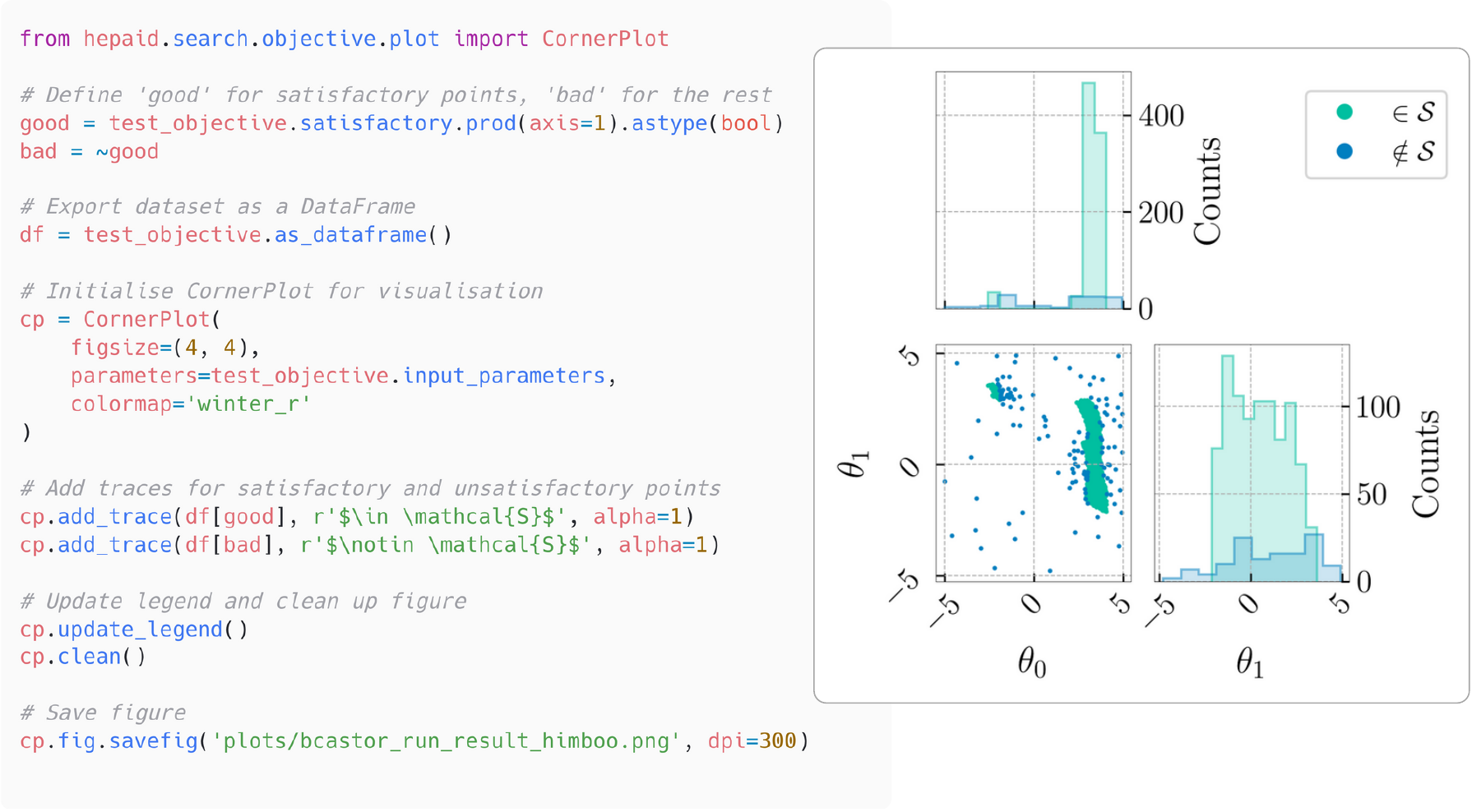}
    \caption{Code snippet that demonstrates the use of the {\tt CornerPlot} utility to visualise data points classified as satisfactory or unsatisfactory. The dataset is filtered using the {\tt Objective.satisfactory} attribute, and traces are added to distinguish between the two classifications.}
    \label{fig:corner_plot_code}
\end{figure}

To provide a practical overview of how to use various functionalities of \hepaid, we utilise the
test function introduced in \cite{madiaz2024}. This is a two-dimensional, double-objective
function comprising both uni-modal and multi-modal objectives, with partially overlapping
constraints for each objective that defines the satisfactory region. The ground truth satisfactory
region is shown in Figure \ref{fig:test_truth}. The general pipeline for using a parameter scan
with \hepaid\ is illustrated in the example code in Figure \ref{fig:quick1}.
When the search method is executed in the command line, a \texttt{progress} bar displays key
metrics related to the search. In Figure \ref{fig:quick1}, which demonstrates the b-CASTOR method,
the progress bar will track the success rate, total points sampled, valid points count, and
satisfactory points count. It also provides the current iteration and the parameter $r$, defining
the radius of the neighbourhood around each parameter vector. Further details on this algorithm are provided in Section \ref{sec:paramscans}. The dataset generated from the search is stored in
\texttt{test\_objective.dataset}. This dataset can be visualised using a corner plot, a technique
that displays pairwise correlations between multiple variables alongside their marginal distributions. In
this context, the plot illustrates the various dimensions of the search space and potentially the
objective dimensions. The corner plot can be generated using the utility module
\texttt{hepaid.search.objective.plot}, as demonstrated in Figure \ref{fig:corner_plot_code}.

\begin{figure}[h!]
    \centering
    \includegraphics[width=0.95\textwidth]{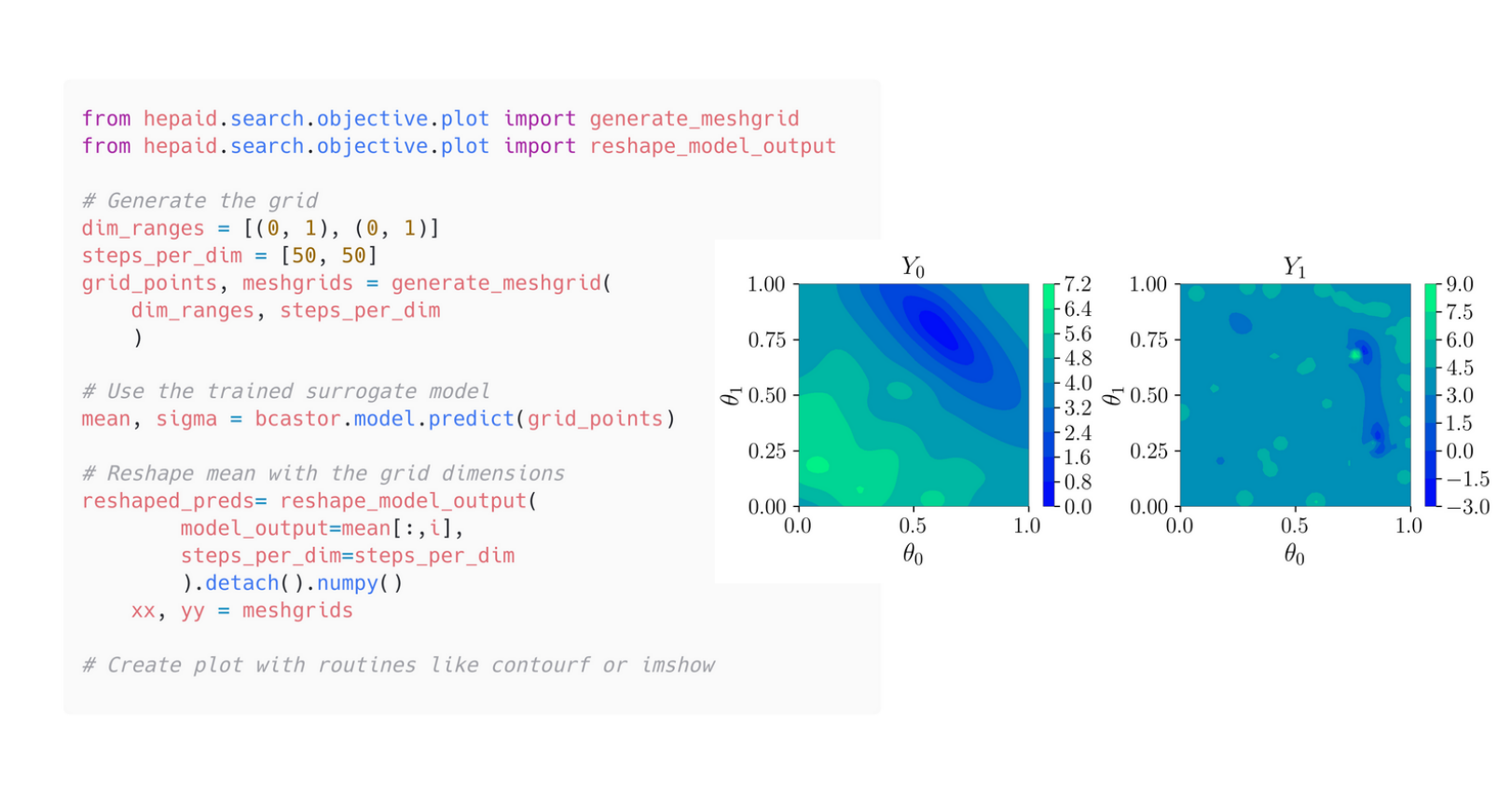}
    \caption{Example code for the application of the \texttt{generate\_meshgrid} and
    \texttt{reshape\_model\_output} routines from the \texttt{hepaid.search.objective.plot}
    module, as well as the evaluation of the GP surrogate model
    \texttt{b-CASTOR.model} on the generated grid, illustrating the fitted model predictions
    based on the current dataset.}
    \label{fig:predictions_him_boo_b-CASTOR}
\end{figure}

Every ML-based active PS method which is used retains a copy of the fitted surrogate model in the
\texttt{method.method} attribute. This model can be accessed after completing the search and will
correspond to the model used in the latest iteration of the search loop. Figure
\ref{fig:predictions_him_boo_b-CASTOR} demonstrates how to use routines from the \texttt{.plot}
module, e.g.,  \texttt{generate\_meshgrid} and \texttt{reshape\_model\_output}, and to create filled contour plots that visualise the objective approximations generated by the
surrogate model fitted to the current dataset.

The \hepaid\ framework currently includes four main PS methods: MCMC-MH \cite{Metropolis:1953am,Hastings:1970aa}, b-CASTOR
\cite{madiaz2024}, CAS \cite{malkomes2021beyond}, and MLScan \cite{ren2019exploring}. (The details
of each algorithm are described in Section \ref{sec:paramscans}.) In short, CAS is an AS approach that uses a GP as a surrogate model. Its search policy aims to
discover the entire satisfactory region in the parameter space where the objectives meet specific
constraints, while also ensuring sample diversity within this set. Here, b-CASTOR is a batched variant
of the CAS algorithm, offering faster discovery of the satisfactory region and a denser filling
of the set. MCMC-MH is an algorithm used to sample a complex target distribution. In the
context of parameter scanning, the latter represents the satisfactory region and is sampled
by defining a likelihood over the objectives. Finally, the MLScan method fits within the active
search paradigm, employing a fully connected NN  as the surrogate model. The
input and output dimensions of this network are determined by the search space and the
objectives, respectively. The MLScan policy involves performing rejection sampling over a
likelihood, which is computed using the surrogate NN model.

\begin{figure}
    \centering
    \includegraphics[width=0.8\linewidth]{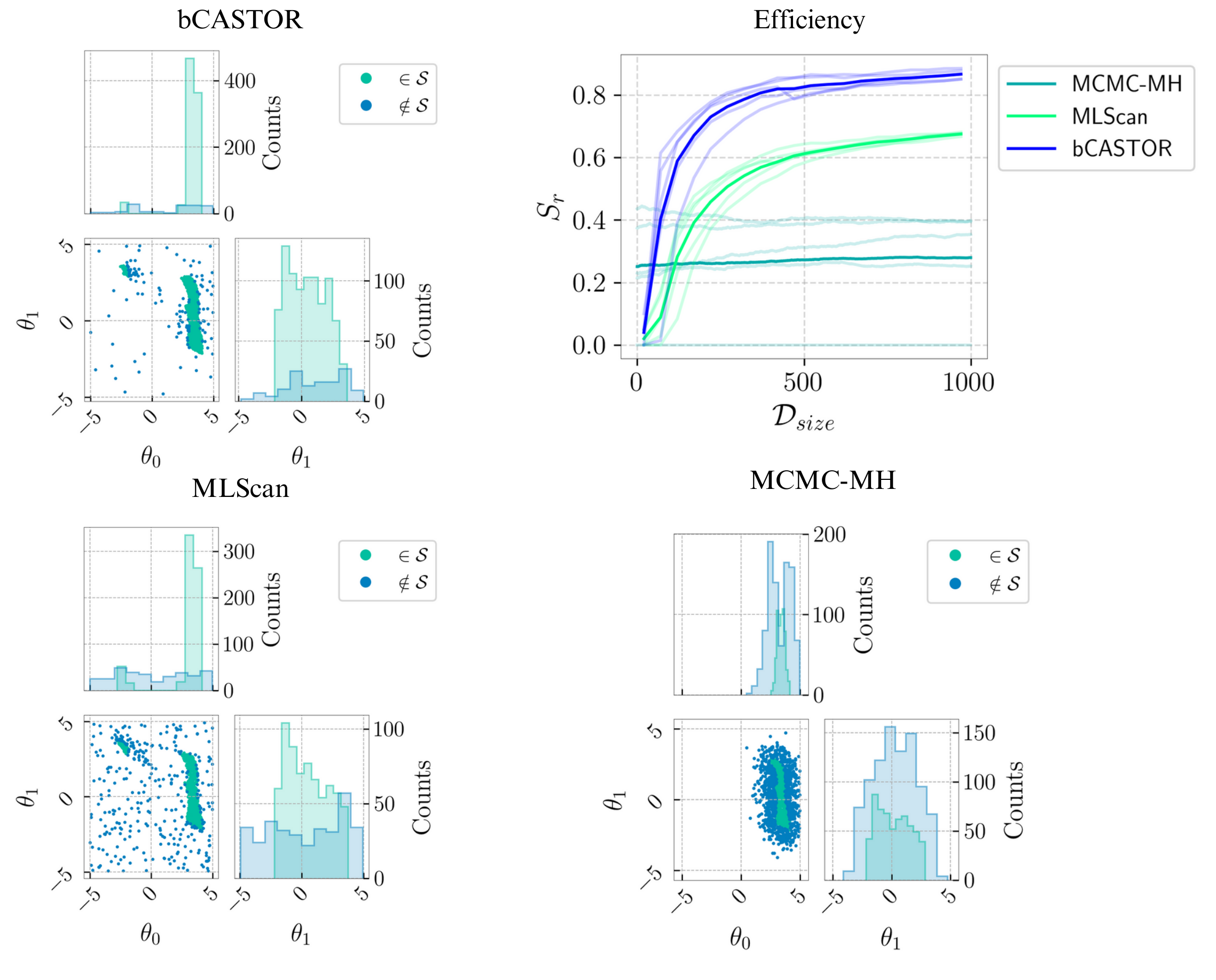}
    \caption{Performance comparison of b-CASTOR, MLScan, and MCMC-MH for a two-dimensional,
    double-objective test function. The mean is displayed for five runs for b-CASTOR as well as MLScan,
    and ten runs for MCMC-MH. The top-right plot displays the efficiency measured as the ratio of
    satisfactory to total points, $\mathcal{S}_r$, as a function of the dataset size,
    $\mathcal{D_{size}}$. The top-left plot shows the corner plot for b-CASTOR, while the
    bottom-left and bottom-right plots present the corner plots for MLScan and MCMC-MH,
    respectively.}
    \label{fig:performance_himboo_all}
\end{figure}

The flexible structure of the \hepaid\ library allows users to switch between methods seamlessly by following
the procedure displayed in Figure \ref{fig:quick1}. This flexibility facilitates straightforward
performance comparisons between different algorithms, as demonstrated in Figure
\ref{fig:performance_himboo_all}. The figure presents the efficiency, measured by the ratio of
satisfactory to total points $\mathcal{S}_r$, as a function of dataset size $\mathcal{D}_{size}$
for b-CASTOR, MLScan, and MCMC-MH on a two-dimensional, double-objective test function. The
search for each algorithm is conducted over five independent runs for b-CASTOR and MLScan, and ten
runs for MCMC-MH. Figure \ref{fig:performance_himboo_all} highlights the superior sample efficiency
of b-CASTOR compared to the other two methods; however, MLScan exhibits greater exploration within
the parameter space, showing robustness for mode discovery. MCMC-MH demonstrates an
$\mathcal{S}_r$ of zero in some runs due to suboptimal random starting points, highlighting a key
weakness of this algorithm\footnote{This analysis can be replicated with the code examples available in \href{https://github.com/mjadiaz/hepaid_tutorials}{github.com/mjadiaz/hepaid\_tutorials}.}. This comparison illustrates the importance of selecting a PS method based on user needs, as each algorithm has distinct strengths that can be
used depending on the requirements of the specific use case.

\subsection{Searching for BSM Physics}
\begin{figure}
    \centering
    \includegraphics[width=0.8\textwidth]{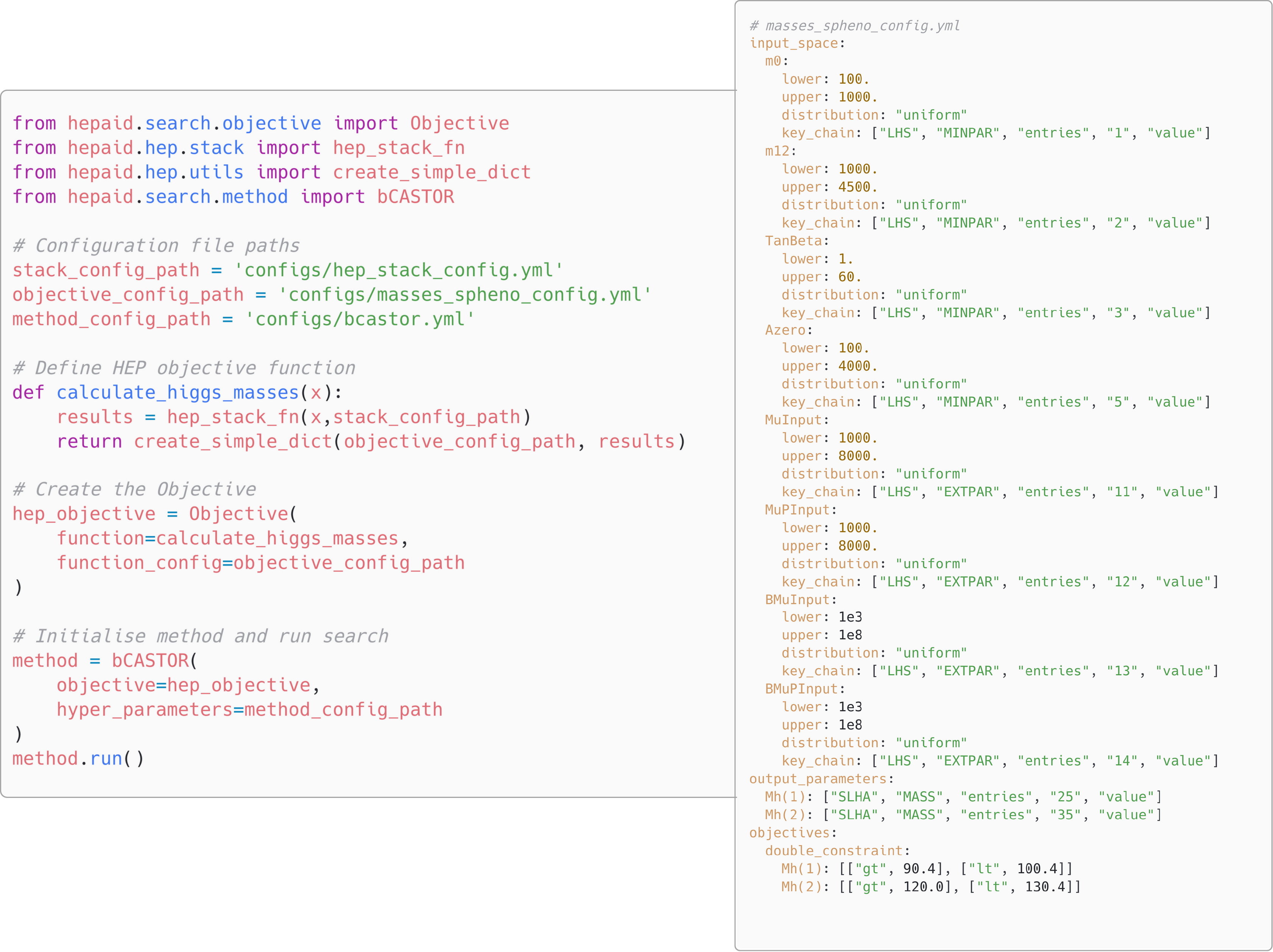}
    \caption{Code example demonstrating the definition of the objective function,
    \texttt{calculate\_higgs\_masses}, and the initialisation of the Objective constructor. Here,
    \texttt{create\_simple\_dict} resolves the key chains in the nested \texttt{result}
    dictionary, retaining only the objective keys and their corresponding values. The search
    configuration file for HEP-stack is also provided, showing the search space configuration, the
    objectives and their constraints.}
    \label{fig:higgs_masses_bcasor_code}
\end{figure}

To perform parameter searches in BSM phenomenology studies, the HEP-stack software must be
installed separately. Therefore \hepaid\ provides a command-line utility for installing the specific HEP
software used in \cite{madiaz2024}. This setup aims to facilitate reproducible phenomenological
scans by sharing BSM spectrum files. By running the command \texttt{hepaid
install-HEP-stack-cli} in the directory containing the SPheno and Universal FeynRules Output (UFO) \cite{Darme:2023hni}  files, \hepaid\ will
install the required software with the appropriate versions\footnote{The \href{https://github.com/mjadiaz/hepaid_tutorials}{companion repository}
provides the  UFO  and SPheno files for the $(B-L)\text{SSM}$ model, enabling the replication of
the example study.}. In this manual, we demonstrate how to conduct a parameter scan within the
$(B-L) \mathrm{SSM}$ to identify Higgs mass values that
can explain an experimental signal around $95 \ \mathrm{GeV}$ in terms of a BSM Higgs boson, alongside the $125 \
\mathrm{GeV}$ SM-like Higgs state. This signal is supported by multiple experimental analyses searching for new
Higgs bosons, including a di-photon ($\gamma \gamma$) excess observed by CMS \cite{CMS:2023yay},
a di-tau ($\tau^+ \tau^-$) excess also reported by CMS \cite{CMS-PAS-HIG-21-001}, and a $b
\bar{b}$ excess detected by LEP \cite{LEPWorkingGroupforHiggsbosonsearches:2003ing}. The search
space in this case is defined by
\begin{equation}\label{eq:muaa_search_space}
\mathcal{X} = \left\{ x \in \mathbb{R}^8 : x = \left( M_0,M_{1/2}, \tan \beta, A_0,\mu, \mu^\prime,B_\mu,B_{\mu^\prime} \right) \right\}.
\end{equation}
For demonstration purposes, we simplify the objective space to include only the masses of the two lightest Higgs particles in the $(B-L)$SSM, i.e., the output space is defined by
\begin{equation}\label{eq:objective_space_aa}
\mathcal{Y} = \left\{ y \in \mathbb{R}^2 : y = \left( m_{h^\prime},m_{h^{\mathrm{SM}}}\right) \right\}.
\end{equation}
Then the task for the search method is to find parameter space points that satisfy the constraints
\begin{equation}\label{eq:muaa_constraints}
  \boldsymbol{\tau} =
    \begin{cases}
      m_{h^{\mathrm{SM}}} &   = 125 \pm \delta m ~{\rm GeV},\\
      m_{h^\prime} &  = 95 \pm \delta m ~{\rm GeV},\\
    \end{cases}       
\end{equation}
where $\delta m$ is a user defined mass window. In this case we will consider $\delta m = 5~{\rm
GeV}$. Figure \ref{fig:higgs_masses_bcasor_code} illustrates the use of the b-CASTOR algorithm to
conduct a search within the $(B-L)$SSM. The objective function must be defined: in this
example, we use the \texttt{hep\_stack\_fn} utility function, which enables the use of a
HEP-stack object for evaluating a single parameter space point. The result is returned as a
nested Python dictionary containing, in this instance, the input and output files for SPheno. To
generate the required output from the objective function, we use the
\texttt{create\_simple\_dict} utility function, which queries the result dictionary as shown
in Figure \ref{fig:using_hep_stack} but in an automated manner. This function takes the list of
keys from the \texttt{masses\_spheno\_config.yml} configuration file and returns only the
specified keys and their values for the input and output parameters defined in the configuration
file. The configuration format is designed to be compatible with the \texttt{Objective} object
constructor by including the \texttt{key\_chain} and \texttt{output\_parameters} elements.

\begin{figure}
    \centering
    \includegraphics[width=0.7\linewidth]{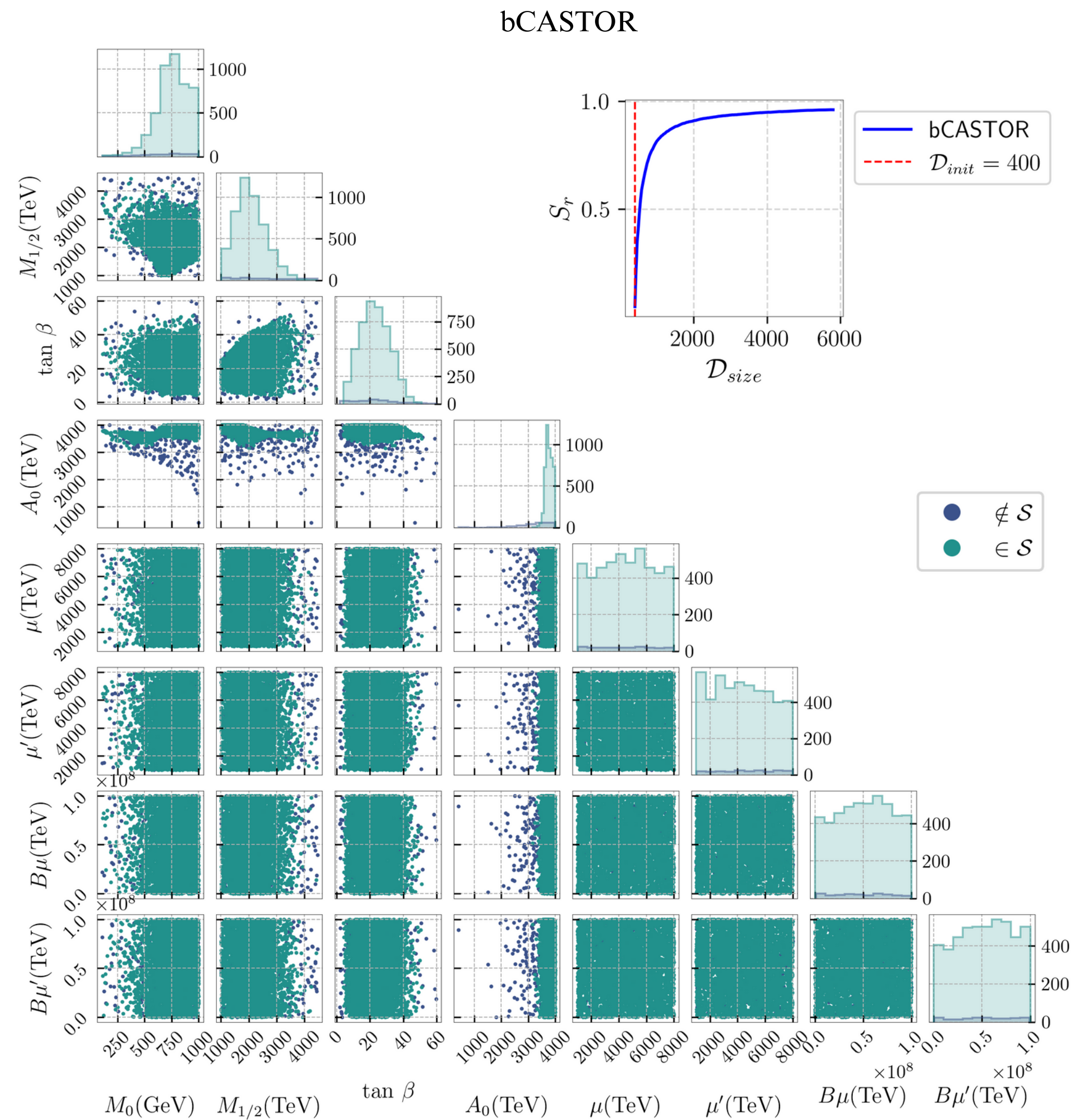}
    \caption{Corner plot visualisation of point distributions within and outside the ${\cal S}$ region. The top-right plot shows the b-CASTOR efficiency, with nearly 95\% of points meeting satisfactory constraints. The red dashed line shows the initial number of points for the search.}
    \label{fig:higgs_masses_bcasor_corner}
\end{figure}

The results can be visualised using the \texttt{CornerPlot} class of \hepaid, as shown in Figure
\ref{fig:higgs_masses_bcasor_corner}, which displays the distribution of points within and
outside the ${\cal S}$ region. In this case, the two classes of points are not clearly separable, as the
distributions of satisfactory and non-satisfactory points overlap, as shown in the marginal
histograms in the corner plots. However, the number of non-satisfactory points is relatively
small, given that b-CASTOR demonstrates nearly 95\% efficiency in the top-right plot. The search
was conducted with an initial dataset of 400 points. The configuration file for b-CASTOR used in
this example is shown in Figure 3.

\section{HEP Module}\label{sec:hepmodule}

The HEP module in \hepaid, located in \texttt{hepaid.hep}, provides infrastructure for managing and
executing HEP software. As mentioned, for phenomenological analysis, a collection of HEP tools, referred to as
a HEP-stack, is typically required to run sequentially for a single parameter space point of the
BSM scenario under study. In fact, \hepaid\ aims to simplify the setup and execution of a
HEP-stack by abstracting away the details of each tool initialisation and execution. It thus allows
the user to perform HEP phenomenology analyses in a simple manner from a Python script or a
Jupyter notebook. 

Currently, part of the SARAH \cite{Staub:2008uz, Staub_2014} family of programs is implemented in
\hepaid. SARAH is a Mathematica package for both Supersymmetric and non-Supersymmetric model
building. SARAH allows the generation of SPheno and UFO model files for a specific BSM model.
SPheno is a model spectrum generator that, given an input SLHA file with a particular parameter
space point, calculates loop-corrected masses, couplings, BRs, decay widths, and
even flavour observables \cite{Porod2003SPhenoAP, Porod_2012}, and write the output in
SLHA format. Using the UFO files generated by
SARAH, MG \cite{Alwall:2014hca} can utilize the SLHA output file from SPheno, known
internally as the \texttt{param\_card.dat}, to calculate process cross-sections and generate 
events. Furthermore, access to the HB \cite{Bechtle_2010} and HS \cite{Bechtle_2014}
programs are also implemented. HB and HS are used for experimental testing of model parameter
space points. HB compares existing exclusion limits from Higgs searches with the model
predictions of the Higgs sector, generating an upper limit to a corresponding signal
cross-section prediction. Therefore, with HB, we can check whether a given model, whose spectrum
is evaluated at a particular configuration, is excluded at the 95\% Confidence Level (C.L.) by
existing Higgs boson searches. In contrast, HS tests the model prediction of a Higgs sector with
an arbitrary number of Higgs bosons against the properties of the observed state as measured by
the LHC experiments ATLAS \cite{atlas2012} and CMS \cite{cms2012} in 2012 (and thereafter).

\subsection{HEP-stack}

\begin{figure}
    \centering
    \includegraphics[width=0.6\linewidth]{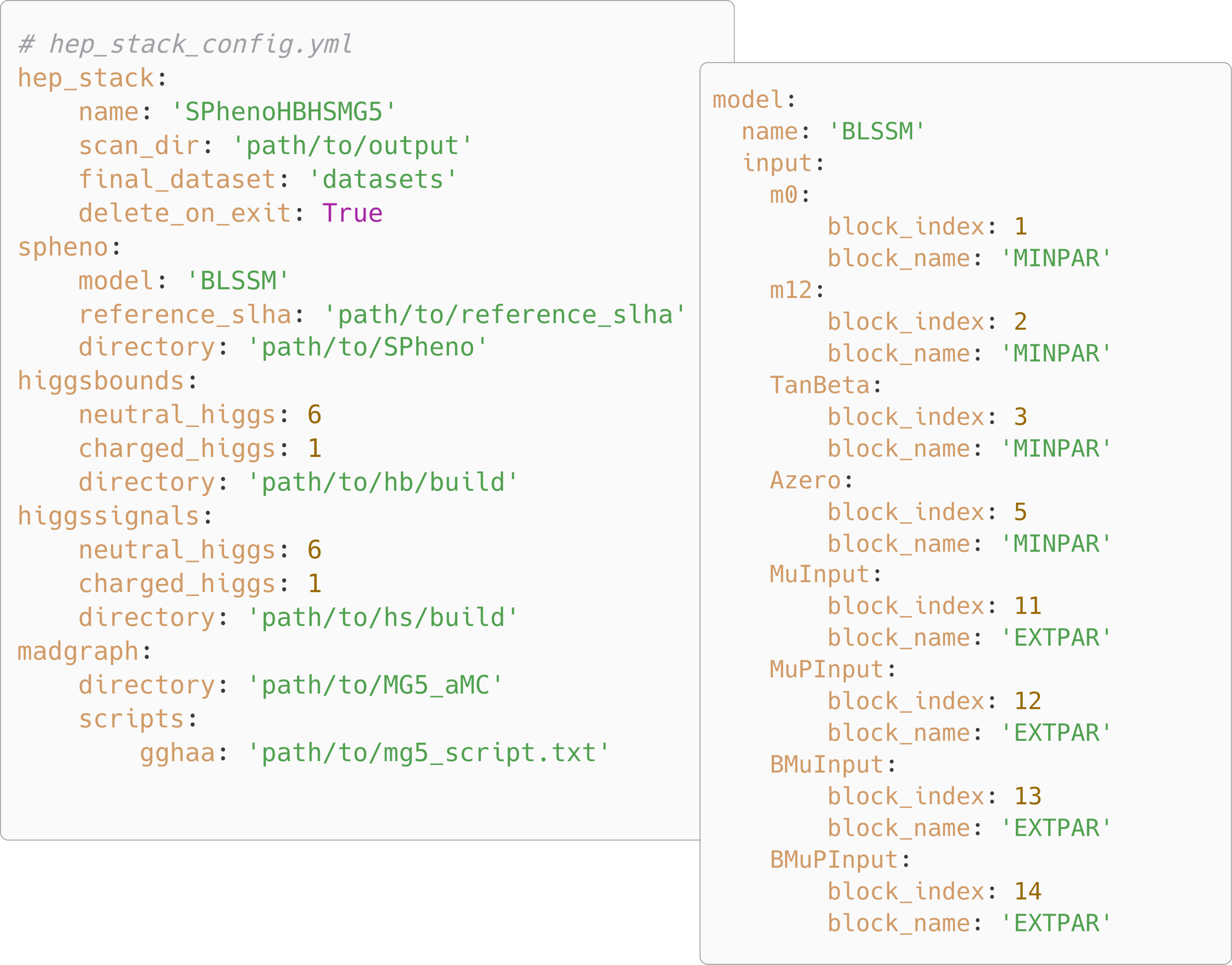}
    \caption{Configuration file for the HEP-stack named \texttt{SPhenoHBHSMG5}. The file includes all
    relevant information for the HEP tools, and on the right, details about the parameters needed to
    conduct a search using SLHA block information.}
    \label{fig:hep_stack_config_yml}
\end{figure}

Every {HEP-stack} is implemented as a Python class and stored in the \texttt{HEPSTACK} dictionary. Each
{HEP-stack} needs to be initialised with a pre-defined configuration file. This configuration file
contains information about the input parameters that will undergo a parameter scan and the
necessary directories for the HEP tools used in the scan. For instance, the {HEP-stack} composed by the 
SPheno-HB-HS-MG sequence can be accessed as shown in Figure \ref{fig:using_hep_stack}.  For this {HEP-stack}, a configuration file is shown in Figure \ref{fig:hep_stack_config_yml}.
With the {HEP-stack} initialised, we can define a Benchmark Point (BP)  within the search space specified
in Eq.~\eqref{eq:muaa_search_space}. We then execute the {HEP-stack} using the \texttt{.sample(x)}
method, as shown in Figure \ref{fig:using_hep_stack}. The result includes all the input and
output information from the four HEP tools used in the HEP-stack, presented as a Python dictionary.
To obtain the objective parameters within $\mathcal{Y}$, defined in Eq. ~\eqref{eq:objective_space_aa}, we can simply query the result dictionary using their
corresponding chain of keys, as illustrated in Figure \ref{fig:using_hep_stack}.

\subsection{HEP Tools}\label{sec:heptools}
The \hepaid\ package implements the \texttt{hepaid.hep.tools} module, which provides classes and
methods for managing high-energy physics tools such as SPheno, MG, HB, and
HS. This module offers a simplified interface for running computations with minimal
boilerplate code. It also handles the automatic management of input and output files, including
the creation of necessary directories, and parses tool-specific output formats into structured
Python objects for further analysis.

\begin{figure}
    \centering
    \includegraphics[width=0.7\linewidth]{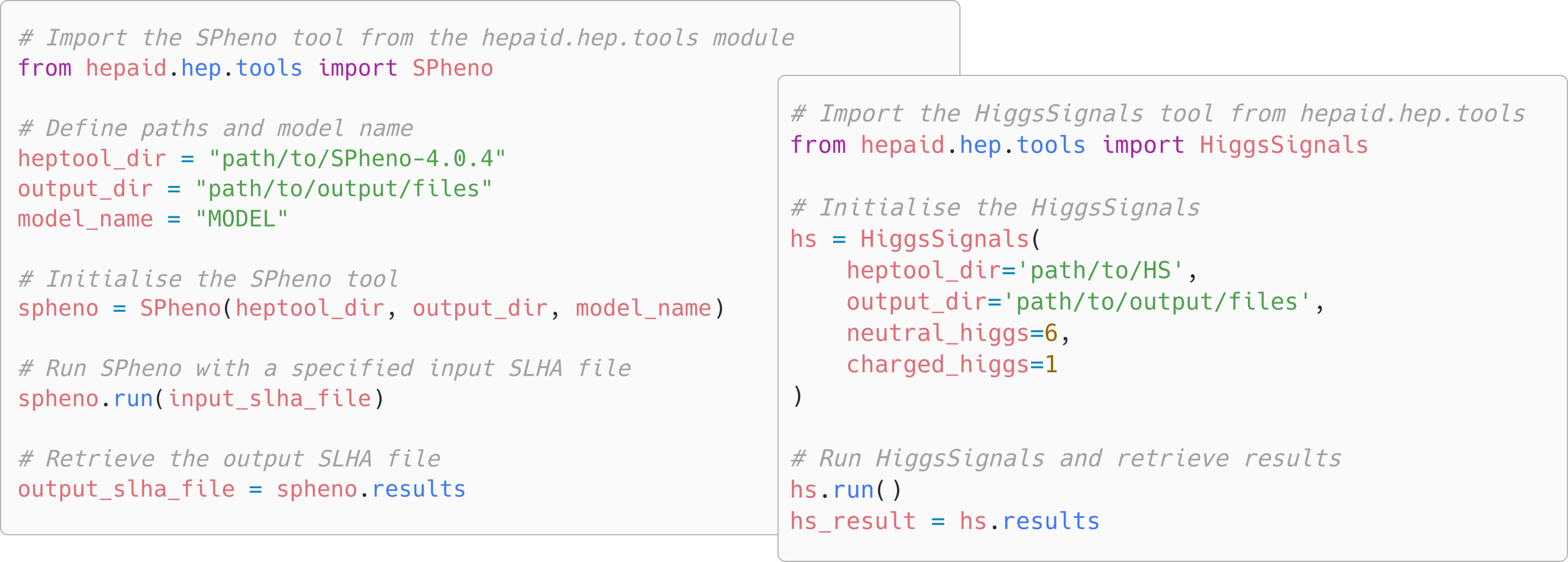}
    \caption{Using SPheno and \texttt{HiggsSignals}: The parameter \texttt{heptool\_dir} corresponds to the
    path to SPheno's directory, \texttt{output\_dir} corresponds to the directory where the output will
    be stored, and \texttt{model\_name} is the name of the previously compiled model. The process 
    is analogous for \texttt{HiggsSignals}, here \texttt{heptool\_dir} refers to the directory containing the build.}
    \label{fig:using_spheno_hs}
\end{figure}

The classes defined within this module corresponds to a specific tool. The structure of each tool
class is defined by the \texttt{BaseTool} class which includes \texttt{run()} and
\texttt{results()} methods. As an example, Figure \ref{fig:using_spheno_hs} demonstrates how
SPheno can be initialised and run in a straightforward manner. HB and HS are also implemented in \hepaid, as mentioned\footnote{The two programs have recently been 
incorporated into the 
HiggsTools distribution \cite{Bahl:2022igd}, which would also be easily embedded in \hepaid.}. The HB and HS tools can be used as shown in Figure 
\ref{fig:using_spheno_hs}, following the same structure as SPheno. Lastly, the
MG HEP tool is implemented to run in script mode. In this case, the
\texttt{run(mg5\_script\_path)} method takes the path to the script as input. The
\texttt{results("path/to/process/output")} method finally takes the path of the MG output generated
by the command \texttt{output example\_process}.

\subsection{Reading and Writing SLHA Files}

The basis for input/ouput  in the HEP module in \hepaid\ is the SLHA format
\cite{Skands:2003cj,Allanach:2008qq}. Libraries for manipulating SLHA files exist in 
literature already. The Pyslha \cite{buckley2015pyslhapythonicinterfacesusy} library is a
well-established and widely used tool for SLHA file manipulation. Pyslha includes functions for
calculating spectrum properties, making it a comprehensive solution for handling SLHA files in
various contexts. Additionally, it offers capabilities for generating mass spectrum plots in
various formats through the \texttt{slhaplot} script. Another notable library is  \cite{Staub_2019_xslha},
which efficiently manages both individual SLHA files and directories containing multiple files,
with a particular emphasis on reading speed optimisation. This library allows for efficient data
filtering through Targeted Data Extraction when specific blocks and entries are known.
Furthermore, xSLHA (a Python parser for files written in the SLHA format) can process large files where spectra are separated by specific keywords.

\begin{figure}[ht]  
    \centering
    \includegraphics[width=0.7\linewidth]{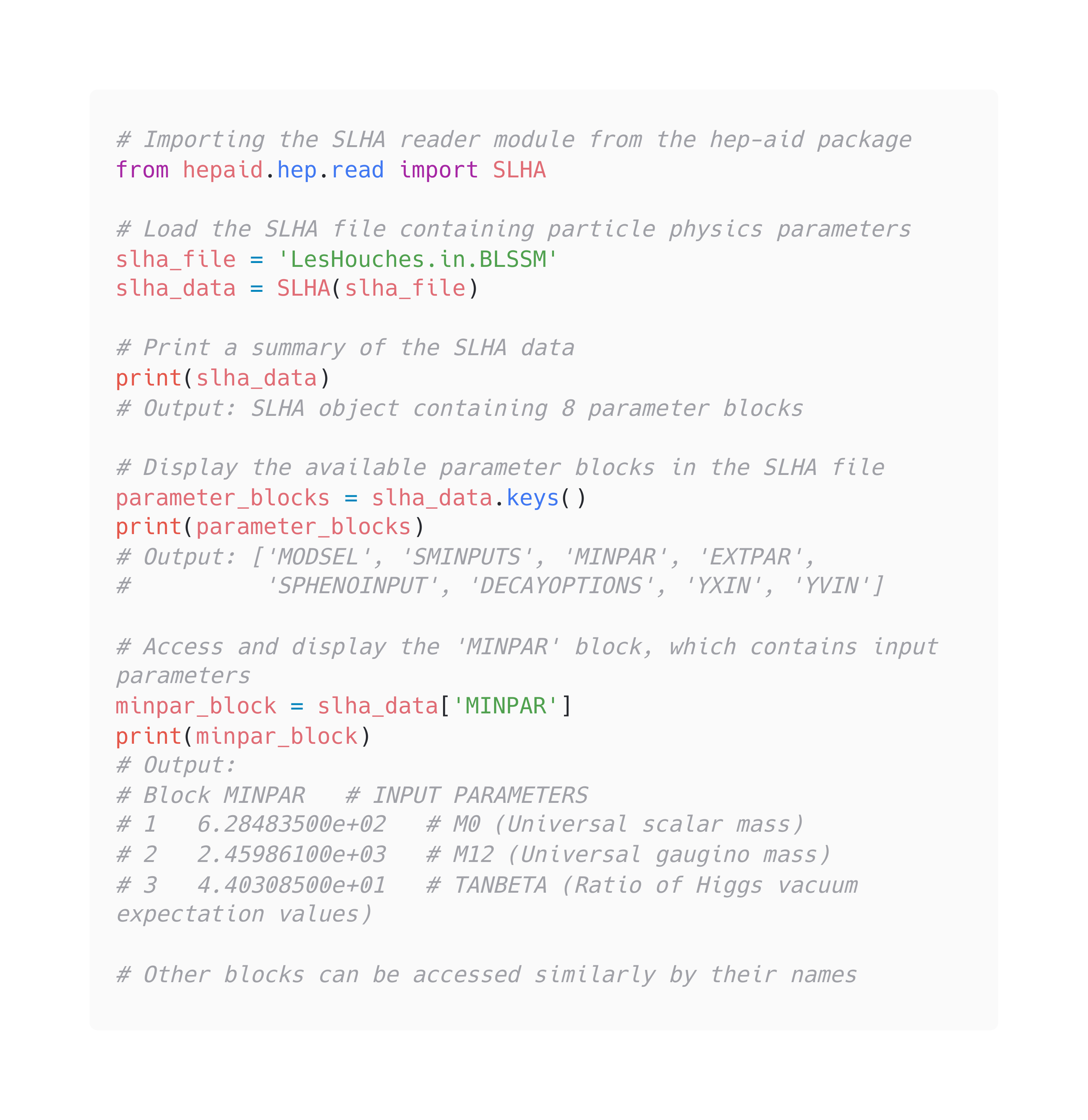}
    \caption{Using the SLHA class for the default input SLHA file of the $(B-L)$SSM. Blocks and entries exhibit dictionary-like behaviour and can be exported as a Python dictionary for efficient storage and communication with other modules. }
    \label{fig:using_slha}
\end{figure}

However, in \hepaid\ implements its own SLHA, {\tt module.lass}, which treats SLHA files like Python
dictionaries, allowing users to access blocks and entries with familiar syntax. This simplifies
data extraction and manipulation without the need for a specialised Application Programming Interface (API). The \texttt{SLHA}  class
retains file comments, preserving valuable metadata, explanations, and references essential for
reproducibility and context. It supports the \texttt{DECAY1L} and \texttt{HiggsTools} blocks, with a
focus on SPheno input files, which include a unique \texttt{"on/off"} configurations. An example on
how the \texttt{SLHA} module is used is displayed in Figure  
\ref{fig:using_slha}. The main objective of creating such a module is to support the construction of large spectrum datasets, by providing functionalities for exporting
SLHA files as nested Python dictionaries, facilitating integration with JavaScript Object Notation  (JSON) and enabling the
storage of multiple SLHA files within a single zipped JSON file. 

Under the hood, every line representing data in a block is structured with a
\texttt{BlockLineSLHA} class which stores the numerical entries, values, comments, and line
category. Note that entries are defined as everything apart from the value of each line. The
blocks are represented with a \texttt{BlockSLHA} class, which stores each line, the block header
and additional information in this header. Lastly, the \texttt{SLHA} extracts automatically the
information of an SLHA file and organises it with \texttt{BlockSLHA} and \texttt{BlockLineSLHA}.
This modular design lets users customise the library  behaviour or add support for new SLHA
blocks, as needed. The \texttt{hepaid.hep.read} module also includes utilities for reading results
from HB, HS, and single MG process generation, to extract the cross-section
and the number of events. These routines are used in the \texttt{hepaid.hep.tools} module, where each
HEP tool includes its results by calling the corresponding \texttt{hepaid.hep.read} routine. As
an example, Figure \ref{fig:using_spheno_hs} shows how SPheno returns the resulting SLHA file by
internally calling the \texttt{SLHA} class in \texttt{hepaid.hep.read}.

\section{Search Module}\label{sec:methods}
In \hepaid, a Parameter Scan Algorithm is implemented as an  AS process. AS is an
iterative search methodology that utilises existing evaluations of an \textit{objective function}
$\mathcal{H}(\mathbf{x})$ to identify suitable points to sample within a required category. In this
context, the required category corresponds to parameter values $\mathbf{x}$ for which the
corresponding observables $\mathbf{y} = \mathcal{H}(\mathbf{x})$ satisfy a specified set of
constraints $\boldsymbol{\tau}$. Therefore, an AS method consists of three main components: the
\textit{objective function}, the \textit{search policy}, and the \textit{surrogate model}. At
each iteration $t$, a dataset $\mathcal{D}_t$ is updated with new evaluations from the
\textit{objective function}. The \textit{surrogate model} is then made to fit this dataset to
refine its approximations and guide the \textit{search policy} in proposing new candidate points
within the satisfactory region. Therefore, technically \hepaid\ comprises three core components. 
In this section, we describe the main components, modules and additional utility functions that
\hepaid\ implements to utilise, implement and develop parameter scan methods.

\subsection{Objective}

The interface between an objective function and a parameter scan method is managed by the
\texttt{Objective} class. This class handles an external multi-objective function, where the
objective function is treated as a black-box. The objective function is defined as a
mapping of an $n$-dimensional input vector $\mathbf{x}$ to an $m$-dimensional output vector
$\mathbf{y}$, with $m$ representing the number of objectives. The objective function is expressed as $\mathbf{y} = \mathcal{H}(\mathbf{x})$. Constraints are defined by the vector
$\boldsymbol{\tau}$, and the satisfactory region in the parameter space, denoted as
$\mathcal{S}$, is the set of configurations that meet these
constraints:
\begin{equation}
    \mathcal{S}=\left\{\mathbf{x} \mid \mathbf{y}=\mathcal{H}(\mathbf{x}) \wedge y_i \succeq \tau_i, i=1, \ldots, m\right\}.
\end{equation}

\begin{figure}[!h]
    \centering
    \includegraphics[width=0.7\linewidth]{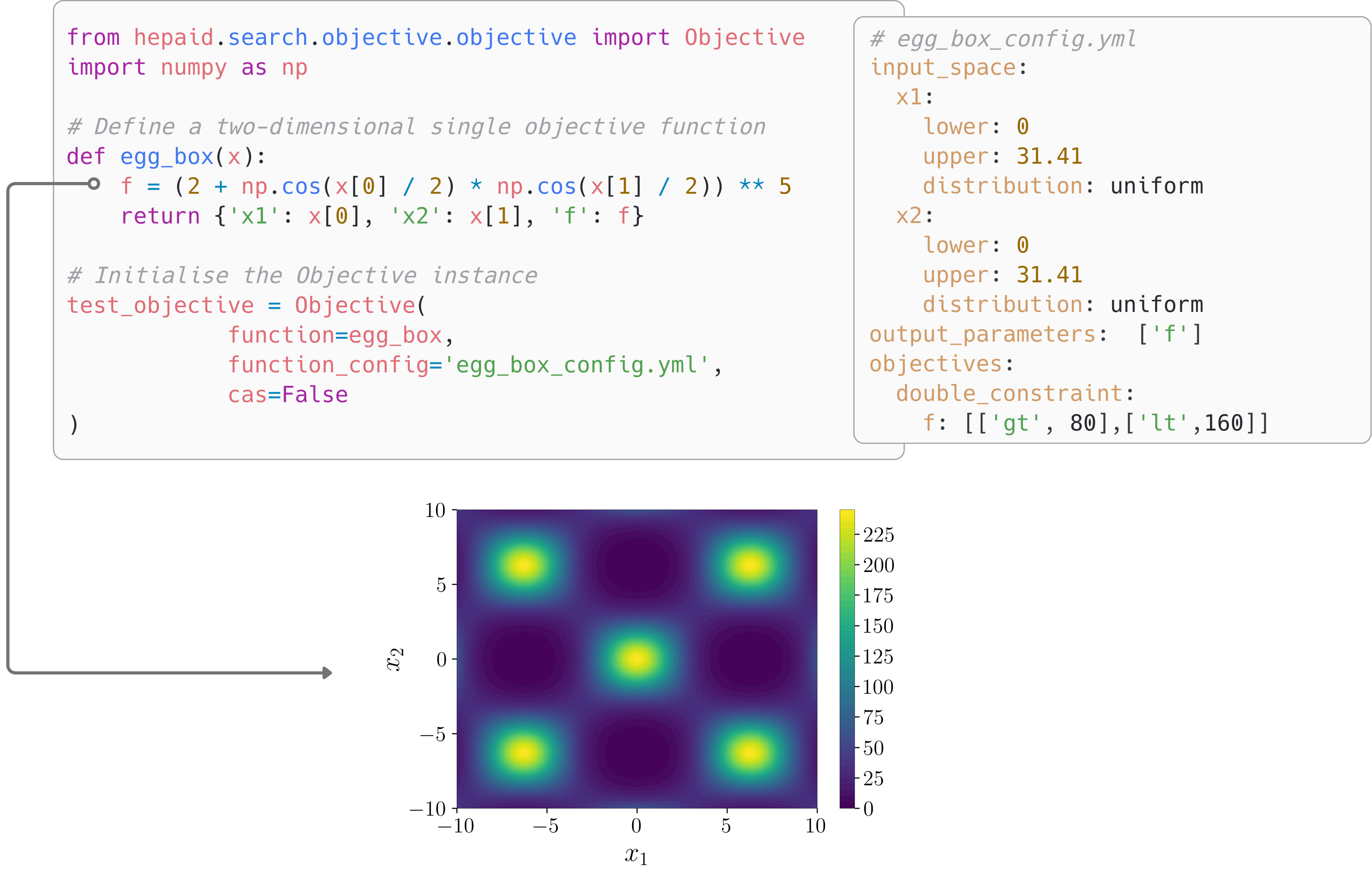}
    \caption{Definition of the Egg Box model as a function, which serves as the objective function to
    initialise the Objective class. The function accepts an array of input parameters and outputs a
    Python dictionary containing all the parameters, both input and output. These parameters and
    their ranges are defined in a configuration file, along with the objective dimensions and their
    constraints. Additionally, the contour plot of the \texttt{egg\_box} function is displayed at the bottom.}

    \label{fig:creating_egg_box}
\end{figure}

By design, the multi-objective function must be defined as a Python function. This function takes
the parameter configuration input $\mathbf{x}$ as an array and outputs a dictionary. The dictionary
includes keys and values that correspond to the names of the input dimensions and the objectives,
along with their respective values. The necessary parameters are defined in the configuration file to initialise the \texttt{Objective} class. This is illustrated in Figure \ref{fig:creating_egg_box} for a
two-dimensional, single-objective Egg Box model function.

The \texttt{Objective} class provides a key feature in \hepaid: the ability to integrate any black-box function. This allows users to input any function, including external software, as long as the function accepts input $\mathbf{x}$ as an array and outputs a vector $\mathbf{y}$. This capability also serves as the mechanism by which the HEP module is integrated into the \texttt{Objective} class.

The \texttt{Objective} object stores the function in its \texttt{Objective.function} attribute and reads the configuration file to create a search space utility using \texttt{scikit-optimize} \cite{head_2021_5565057}. This search space facilitates the normalisation of the internal dataset, stored in \texttt{Objective.dataset}, which is essential for fitting \textit{surrogate models}. The dataset is dynamically updated whenever the function is called through the \texttt{Objective.sample(x)} method, adding the new set ${\mathbf{x}, \mathbf{y}}$ to the internal dataset. The historical \texttt{Objective.dataset} is formatted as a
Python dictionary. However, the \texttt{Objective.x}, \texttt{Objective.X}, and
\texttt{Objective.Y} attributes hold the input data in the original space, the input data in the
normalised space, and the output data in the original space, respectively. The method
\texttt{Objective.as\_dataframe(satisfactory=True)} return the dataset as a Pandas DataFrame
\cite{reback2020pandas}. The \texttt{Objective.satisfactory} attribute generates a boolean matrix, $S_{ij}$, with the same dimensions as \texttt{Objective.Y}. This matrix identifies which elements in \texttt{Objective.Y} satisfy the constraints defined in the configuration file. Consequently, taking the vertical product, $S_j = \prod_i S_{ij}$, of \texttt{Objective.satisfactory} produces a boolean array. This array labels as \texttt{True} the configurations in \texttt{Objective.dataset} that meet the specified constraints.

Lastly, the method \texttt{Objective.save(path)} saves the internal dataset as a compressed JSON
file. Then, the method \texttt{Objective.load(path, process=True)} loads a previously saved JSON
file, updating the internal dataset and all relevant information if the \texttt{process} argument
is set to \texttt{True}. This allows the continuation of the search process with a parameter scan
algorithm or for plotting purposes.

\subsection{Parameter Space Sampling}\label{sec:paramscans}

The \texttt{hepaid.search.methods} module includes all the implemented PS  algorithms.
Currently, these methods are MCMC-MH \cite{Metropolis:1953am,Hastings:1970aa}, b-CASTOR \cite{madiaz2024}, CAS \cite{malkomes2021beyond},
and MLScan \cite{ren2019exploring}. Due to the varying requirements of each parameter scan
algorithm, \hepaid\ implements a base class called \texttt{Method}. This class is designed to be
inherited by PS algorithms and encapsulates the essential functionalities needed for
managing hyperparameters, tracking standard and custom metrics, handling checkpoints, and
managing file directories. Every method that inherits from this base class will receive an
\texttt{Objective} instance and a hyperparameter file as arguments. The file can be provided by the user as a
string path or a \texttt{DictConfig} \cite{omegaconf2019}. If neither is provided, the default
configuration file will be used. The \texttt{Method} class also offers utility methods to save
and load the state of the search process, including the state of the objective function, current
metrics being tracked, and the hyperparameter configuration file. Each instance of the
\texttt{Method} class initialises a \texttt{Metrics} object responsible for tracking standard
metrics related to the search process. These metrics include the total number of points
evaluated, the number of valid points, the number of satisfactory points, the success rate, and
the current iteration counter. Additionally, users can extend the metrics functionality by
defining custom metrics that can be updated during the search process.

There is a defined general structure for the hyperparameter configuration file, as illustrated
in Figure \ref{fig:higgs_masses_bcasor_config}. Each instance of the \texttt{Method} class will
read the following hyperparameters.  The \texttt{run\_name} refers
to the path to the directory where checkpoints and configuration files will be saved. The
\texttt{parallel} parameter is defined as a boolean parameter to enable the parallel evaluation
of specific processes during the search. This works in conjunction with the \texttt{n\_workers}
parameter, which dictates the number of parallel processes based on available resources. For
example, this setup is used for evaluating the objective function of the batch $\boldsymbol{X}^*$
suggested by the policy, by using the function \texttt{batch\_evaluation} located in
\texttt{hepaid.search.objective.utils}. Except for MCMC-MH, search methods require an initial
dataset, which is configured via the \texttt{initial\_dataset} hyperparameter. The
\texttt{initial\_dataset.n\_points} specifies the number of points to be generated, while
\texttt{initial\_dataset.generate} is a boolean indicator for the dataset generation. This is
useful, for instance, when the user wishes to continue from previous checkpoints where the complete initial dataset is not needed. The initial dataset is generated by the function
\texttt{generate\_initial\_dataset}. The \texttt{total\_iteration} parameter sets the total
number of iterations for the search loop. Additional parameters will depend on the specific
algorithm and are explained below.

\begin{table}[h!]
    \centering
    \resizebox{\textwidth}{!}{
    \begin{tabular}{ c c }
    \hline
    \textbf{Hyperparameter} & \textbf{Description} \\ \hline
    \texttt{run\_name} & Directory path for saving checkpoints and configuration files. \\ \hline
    \texttt{parallel} & Boolean parameter to enable parallel evaluation of processes. \\ \hline
    \texttt{n\_workers} & Number of parallel processes based on available resources. \\ \hline
    \texttt{initial\_dataset} & Configures initial dataset required by most search methods. \\ \hline
    \texttt{initial\_dataset.n\_points} & Number of points to generate in the initial dataset. \\ \hline
    \texttt{initial\_dataset.generate} & Specifies whether the initial dataset should be generated. \\ \hline
    \texttt{total\_iteration} & Sets the total number of iterations for the search loop. \\ \hline
    \end{tabular}
    }
    \caption{Hyperparameters for a PS method implemented by inheriting the \texttt{Method} class.}
    \label{tab:base_hp}
    \end{table}

\subsubsection{AS Methods}

As mentioned, 
\hepaid\ implements two AS methods, CAS \cite{malkomes2021beyond}
and the b-CASTOR \cite{madiaz2024} algorithm, which utilises GPs as
surrogate models to approximate each objective function. Herein, GPs provide a probabilistic model where
each point in the search space has an estimated mean, the predicted objective value, and a
standard deviation, uncertainty of the prediction. The Expected Coverage Improvement (ECI)
function is then applied to guide the selection of points to evaluate. ECI operates by defining a
hypersphere around each point in the current dataset and measuring the volume covered by these
hyperspheres across the search space by using the predictions from the GP surrogates. The main
goal is to find a location $\mathbf{x}^*$ in the search space for a new hypersphere that will
maximally increase the coverage of the $\mathcal{S}$ region. At each iteration, a global
optimisation technique is then applied to the ECI function to determine the best new point,
$\mathbf{x}^*$. The resulting dataset from this search process will consist of a diverse set of
points that effectively cover the $\mathcal{S}$ region, with diversity determined by the radius of
each hypersphere, denoted by $r$. This radius $r$ is a fixed hyperparameter within the method.
However, the \hepaid\ implementation of CAS includes a linear decay on the radius, referred to in
the library as the resolution parameter.

Therefore, the additional hyperparameters relevant for CAS are located in 
\texttt{eci.num\_samples}, which specifies the number of sample points for which an individual
hypersphere is approximated, and the \texttt{resolution} block. In this context,
\texttt{resolution.constant\_resolution} allows for switching between a fixed radius or a linear
decay in the radius. The \texttt{resolution.value} sets the value, typically ranging between 0
and 1, as the search space is normalised. The parameters \texttt{resolution.initial} and
\texttt{resolution.final} are used when the linear decay mode is enabled. Additionally,
\texttt{resolution.r\_decay\_steps} specifies the number of iteration steps during which the
decay takes place.

\begin{table}[h!]
\centering
\resizebox{\textwidth}{!}{
    \begin{tabular}{ c c }
    \hline
    \textbf{Hyperparameter} & \textbf{Description} \\ \hline
    \texttt{eci.num\_samples} & Number of sample points for approximating each hypersphere. \\ \hline
    \texttt{resolution.constant\_resolution} & Switches between fixed radius and linear radius decay. \\ \hline
    \texttt{resolution.value} & Sets radius value, typically between 0 and 1 (normalised space). \\ \hline
    \texttt{resolution.initial} & Initial radius value for linear decay. \\ \hline
    \texttt{resolution.final} & Final radius value for linear decay. \\ \hline
    \texttt{resolution.r\_decay\_steps} & Number of steps for radius decay. \\ \hline
    \end{tabular}}
\caption{Hyperparameters used for CAS in \hepaid.}
\label{tab:cas_hp}
\end{table}

CAS operates sequentially, selecting only a single point per iteration to add to the dataset.
This approach leads to relatively slow convergence, as the search progresses incrementally,
expanding the $\mathcal{S}$ region step-by-step to nearby areas. A batched version of this
method, called b-CASTOR, was developed in \cite{madiaz2024}. It uses the same foundational
components as CAS, employing GPs as surrogates and the ECI function to guide the search. However,
different optimisation and sampling methods are applied to the ECI function to generate a batch
of samples for evaluation in the objective function during each iteration. Specifically, b-CASTOR leverages the Tree-structured Parzen Estimator (TPE) to optimise the ECI. Furthermore, a Stochastic Prioritisation (SP) sampling technique is employed to select the batch $\boldsymbol{X}^*$ from the TPE optimisation history. This technique alternates between an exploitation phase -- sampling from high-potential regions as predicted by the \textit{surrogate model} -- and an exploration phase, which uniformly samples the parameter space. This batched approach accelerates the search process by enabling more
diverse and densely populated sample collections in each iteration, thereby improving convergence
speed and coverage of high ECI regions.

The additional hyperparameters required by b-CASTOR include those from CAS related to the
resolution parameter, as well as the following ones. For the batch sampling settings, we have
\texttt{batch\_sampling.tpe\_trials}, which indicates the number of trials for optimising the ECI
function using the TPE algorithm. Next, \texttt{batch\_sampling.rank\_samples} specifies the
number of samples obtained from the historical data of the TPE optimisation using the stochastic
prioritisation technique. This process determines the size of the batch $\boldsymbol{X}^*$. The
\texttt{batch\_sampling.alpha} parameter controls the degree of prioritisation in rank-based
sampling; a higher value prioritises sample efficiency. The interaction between the \texttt{alpha} and \texttt{rank\_samples} hyper-parameters determines the balance between exploration and exploitation in the algorithm. Empirically, setting \texttt{alpha=2} provides a good balance
of efficiency and diversity within the satisfactory region. A higher number of
\texttt{rank\_samples}, i.e., a larger batch size, allows for more chances for exploration.

\begin{figure}
   \centering
    \includegraphics[width=0.6 \textwidth]{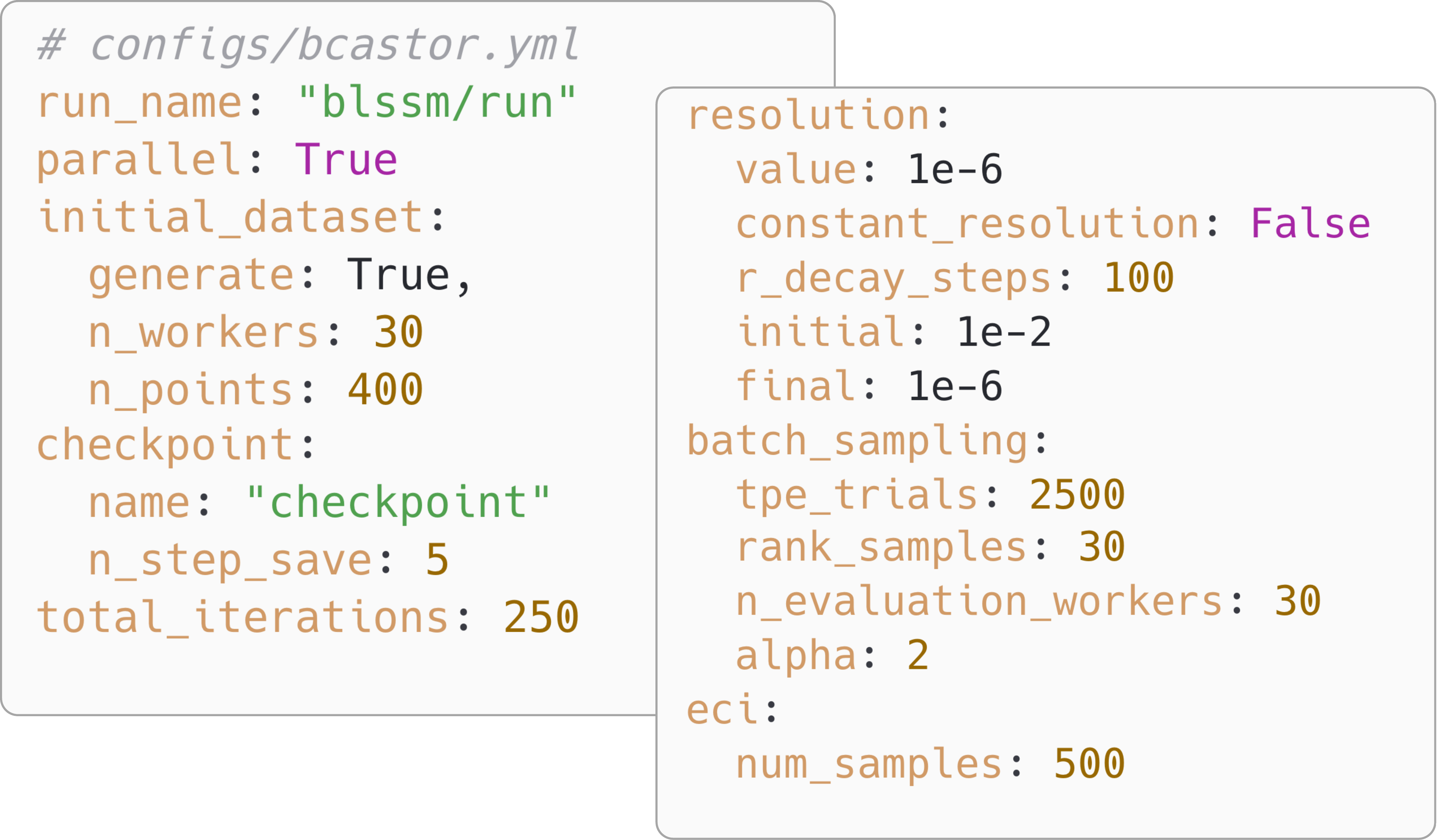}
\caption{Configuration file for the b-CASTOR algorithm. The left side displays the general structure of a
configuration file for a parameter scan implemented in \hepaid. The right side displays the parameters
relevant to the CAS algorithm  in the resolution block, where additional batch sampling
blocks are required for the b-CASTOR algorithm.}
    \label{fig:higgs_masses_bcasor_config}
\end{figure}

\begin{table}[h!]
    \centering
    \resizebox{\textwidth}{!}{
    \begin{tabular}{ c c }
    \hline
    \textbf{Hyperparameter} & \textbf{Description} \\ \hline
    \texttt{batch\_sampling.tpe\_trials} & Number of TPE trials for optimising the ECI function. \\ \hline
    \texttt{batch\_sampling.rank\_samples} & Number of samples obtained through rank based sampling (batch $\boldsymbol{X}^*$ size). \\ \hline
    \texttt{batch\_sampling.alpha} & Prioritisation degree in rank-based sampling. \\ \hline
    \end{tabular}
    }
    \caption{Additional hyperparameters (with respect to the CAS ones) relevant to the b-CASTOR algorithm.}
    \label{tab:b-CASTOR_hp}
    \end{table}
Figure \ref{fig:higgs_masses_bcasor_config} shows the complete configuration file for the b-CASTOR
algorithm, which can also be used for the CAS algorithm. The file is displayed in \texttt{YAML}
format.

The technical implementation for both the CAS and b-CASTOR algorithms is as follows. We used the
ECI acquisition function implementation provided by the BoTorch library
\cite{balandat2020botorch}. Additionally, we integrated GP models, available in
\texttt{hepaid.search.models}, from the GPytorch library \cite{NEURIPS2018_27e8e171}, a repository 
for scalable GP inference built on PyTorch \cite{paszke2019pytorch}. We further utilise
the TPE implementation available in Optuna \cite{optuna_2019}, an open-source hyperparameter
optimisation framework.

\subsubsection{MCMC-MH}
The MCMC-MH algorithm is a method for sampling
from a complex probability distributions \cite{Hogg_2018}. The use of this sampling method and its advanced variants are
well-established in HEP research \cite{albert2024comparison}. The algorithm generates samples that
approximate the target distribution by constructing a Markov chain, which  stationary distribution
is the desired target distribution. The MCMC-MH algorithm starts by selecting an initial state $x_0$
from the state space. During each iteration $t$, a candidate state $x'$ is generated from a
proposal distribution $q(x' \mid x_t)$. The acceptance probability is then computed as
\begin{equation}
\alpha(x', x_t) = \min\left(1, \frac{\pi(x') q(x_t \mid x')}{\pi(x_t) q(x' \mid x_t)}\right), 
\end{equation}
where $\pi(x)$ is the target distribution and $q(x' \mid x_t)$ is the proposal distribution.
A common choice for the proposal distribution in the MCMC-MH algorithm is a Gaussian distribution
centered on the current sample which simplifies the acceptance probability. This proposal is implemented in \hepaid. The proposal variance is
crucial, as a large one can lead to proposals being frequently rejected, which slows down the
algorithm and decreases sample efficiency. Conversely, a low variance can cause the chain to
explore the target distribution inefficiently, potentially preventing the chain from converging
to a stationary distribution. (The variance is also known as the step size or scale parameter.)
A uniform random number $u$ is generated from the interval $[0, 1]$. If $u \leq \alpha(x', x_t)$,
the candidate is accepted by setting $x_{t+1} = x'$, otherwise, the candidate is rejected, and one 
sets $x_{t+1} = x_t$. This iteration process is repeated until a sufficient number of samples have
been generated.

The hyperparameters defined in the configuration file include \texttt{burn\_in}, which defines
the number of initial iterations to discard, allowing the chain to converge to the target
distribution and helping to eliminate dependence on the starting value. The
\texttt{initial\_scale} parameter sets the initial step size for generating proposal states from
the current state. The \texttt{adapt\_frequency} parameter defines how often to adjust the scale
of the proposal distribution during the sampling process, based on the observed acceptance rate.
Lastly, the \texttt{target\_acceptance\_rate} specifies the desired acceptance rate for proposal
moves, guiding the adaptation of the proposal distribution scale and influencing the amount of
exploration of the parameter space while ensuring efficient convergence to the target
distribution.

\begin{table}[h!]
\begin{center}
    \resizebox{0.8\textwidth}{!}{
    \begin{tabular}{ c c }
    \hline
    \textbf{Hyperparameter} & \textbf{Description} \\ \hline
    \texttt{burn\_in} & Number of initial iterations for burn-in. \\ \hline
    \texttt{initial\_scale} & Initial step size for generating proposal states. \\ \hline
    \texttt{adapt\_frequency} & Frequency of adjusting the scale on burn-in. \\ \hline
    \texttt{target\_acceptance\_rate} & Desired acceptance rate for proposals. \\ \hline
    \end{tabular}
    }
    \caption{Hyperparameters relevant for the MCMC-MH algorithm.}
    \label{tab:mcmc_hp}
    \end{center}
    \end{table}

For a PS, the target distribution is the likelihood constructed from the
constraints in the objectives. In \hepaid, this likelihood can be provided as a function. If not,
\hepaid\ will use a default likelihood constructed with sigmoid windows for each objective and its
constraints \cite{madiaz2024},
\begin{equation}
    \mathcal{L}(y_i) = 
      \begin{cases}
         \sigma(y_i, a)&   y_i > a\\
        1 - \sigma(y_i, a) &  y_i < a \\
          \sigma(y_i, a) - \sigma(y_i,b) &    a < y_i < b \\
      \end{cases}       
  \end{equation}
where $y_i$ is an objective and $a,b$ constraints. Here, $\sigma$ is the sigmoid function and is defined as
  \begin{equation}
  \sigma(y,a)=\frac{1}{1+e^{-(y - a) / \epsilon}}, 
  \end{equation}
where $\epsilon$ is a parameter that controls the smoothness of the sigmoid function and $a$ shifts the centre of the sigmoid. Then, the total likelihood for a point $(\mathbf{x},\mathbf{y})$ used for MCMC-MH sampling is defined as
  \begin{equation}
      \mathcal{L}(\mathbf{y}) = \prod_i \mathcal{L}(y_i).   
  \end{equation}

\subsubsection{MLScan}

The Machine Learning Scan method  \cite{ren2019exploring}, termed MLScan in \hepaid\, was
designed for efficient sampling and exploration of parameter spaces using a Machine Learning
surrogate model for the observables, fitting in the description of AS. It uses an MLP Neural Network as a \textit{surrogate model}. The search policy in this case 
is performing rejection sampling over the likelihood, but this time the likelihood is evaluated using 
the predictions from the MLP model. 

The implementation in \hepaid\ begins with an initial dataset, similar to previous methods. In
each iteration, the MLP is trained using the currently available dataset, retaining the model
parameters from previous iterations to perform incremental learning. The rejection sampling
technique is used to generate the batch $\boldsymbol{X}^*$, with the size of this batch
controlled by the hyperparameter \texttt{num\_samples}. The scaling of the acceptance
probability for the rejection sampling method is controlled by the parameter \texttt{m\_factor}.
By adjusting this factor, users can manage the balance between sample quality and the diversity
of generated samples, thus affecting both the efficiency and effectiveness of the sampling
process. The MLScan method also adds extra random samples to encourage exploration, with the
number of samples controlled by the hyperparameter \texttt{extra\_random\_samples}.

\begin{figure}[h!]
    \centering
    \includegraphics[width=0.7\textwidth]{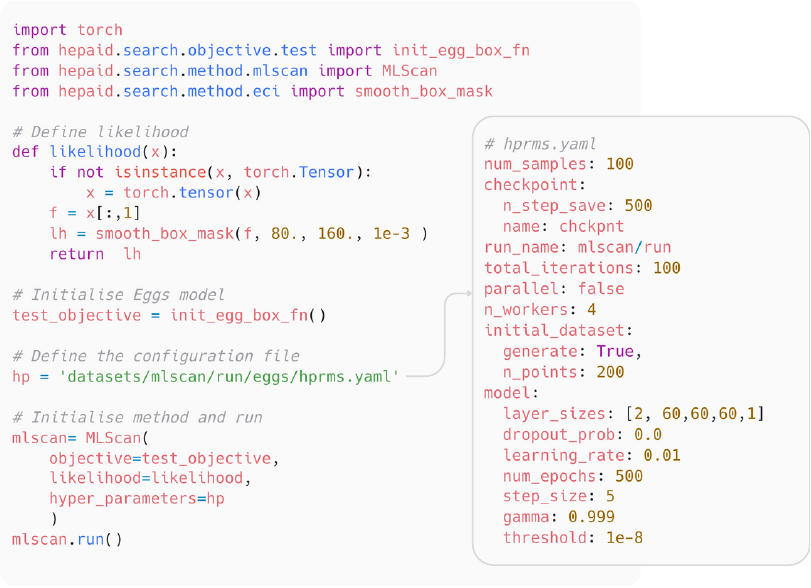}
    \caption{A code example demonstrating the utilisation of the MLScan method, a neural
    network-based sampling method introduced in \cite{ren2019exploring}, is provided.  Users must
    define a likelihood function. Since the Egg Box model is already implemented in \hepaid, it
    can be readily used. Initialising the MLScan method requires a configuration file, as shown
    in the code snippet on the right. }
    \label{fig:eggs_replicate_code}
\end{figure}

For MLScan an additional hyper-parameter block, named \texttt{model\_hyperparameters}, is
required for configuring the MLP's architecture and training. The \texttt{layer\_sizes} parameter
receives a list specifying an input layer, an arbitrary number of hidden layers with their
respective number of neurons, and the output layer neurons. The \texttt{dropout\_prob} indicates
the level of dropout regularisation in each layer. The \texttt{learning\_rate} controls the
magnitude of weight updates during training, while \texttt{num\_epochs} defines the number of
complete passes through the training dataset. Additionally, the \texttt{step\_size} determines
how frequently the learning rate is adjusted, and \texttt{gamma} specifies the decay rate for the
learning rate, allowing it to decrease gradually. Finally, the \texttt{threshold} acts as a
stopping criterion for training when the loss function fall below this value.

\begin{figure}[h!]
    \centering
    \includegraphics[width=0.6\linewidth]{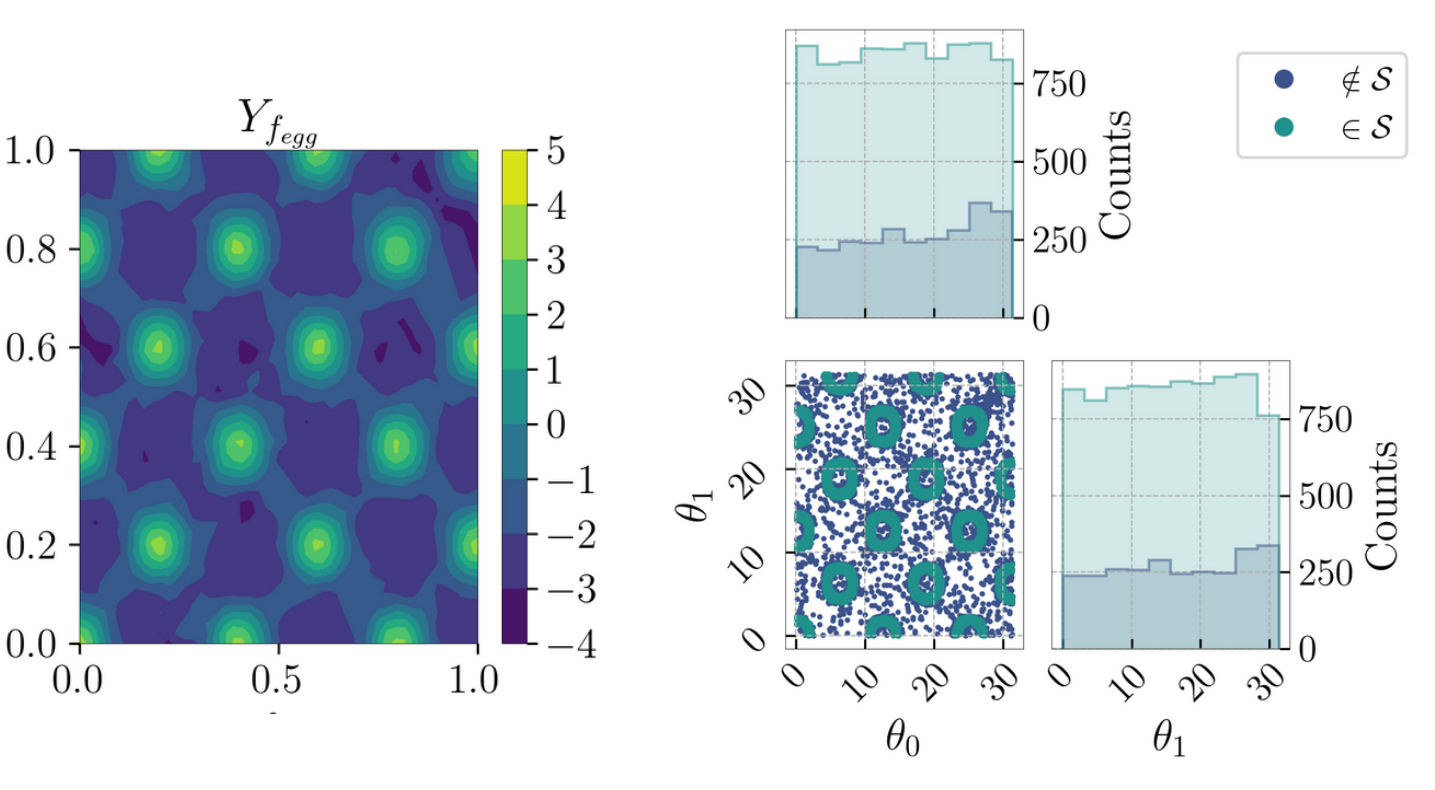}
    \caption{On the left, the predictions across the entire parameter space of the surrogate
    neural network model in the final iteration of the search from Figure
    \ref{fig:eggs_replicate_code} are shown. On the right, the points identified by the MLScan
    method are displayed using the corner plot functionalities of \hepaid.}
    \label{fig:eggs_replicate_results}
\end{figure}

Finally, the \texttt{MLScan} class takes an argument for the likelihood function, which the user
must define, as shown in Figure \ref{fig:eggs_replicate_code}. In this example, \hepaid\ can be
used to replicate the results from \cite{ren2019exploring} using the Egg Box model test function,
which is currently implemented in \hepaid. The results of running the MLScan
search on the Egg Box model are displayed in Figure \ref{fig:eggs_replicate_results}.

\begin{table}[h!]
    \centering
    \resizebox{0.8\textwidth}{!}{
    \begin{tabular}{ c c }
    \hline
    \textbf{Hyperparameter} & \textbf{Description} \\ \hline
    \texttt{num\_samples} & Batch size for rejection sampling. \\ \hline
    \texttt{m\_factor} & Scales acceptance probability in rejection sampling. \\ \hline
    \texttt{extra\_random\_samples} & Number of additional random samples for exploration. \\ \hline 
    \texttt{model\_hyperparameters} & Configures MLP architecture and training. \\ \hline 
    \texttt{layer\_sizes} & Defines neuron counts for MLP layers. \\ \hline
    \texttt{dropout\_prob} & Dropout rate for regularisation. \\ \hline
    \texttt{learning\_rate} & Step size for weight updates. \\ \hline
    \texttt{num\_epochs} & Total training dataset passes. \\ \hline
    \texttt{step\_size} & Interval for adjusting learning rate. \\ \hline
    \texttt{gamma} & Learning rate decay factor. \\ \hline
    \texttt{threshold} & Loss threshold for stopping criterion. \\ \hline
    \end{tabular}
    }
    \caption{Hyperparameters relevant for the MLScan algorithm.}
    \end{table}

\section{Conclusions}\label{sec:summa}
This paper introduced \hepaid, a new Python library designed to facilitate PS
algorithms for BSM phenomenology. The library provides a modular framework that
integrates HEP software, simplifying the implementation and development of
PS algorithms while offering essential functionalities for phenomenological
analysis. Originally the library was created to give researchers simple access to use the b-CASTOR and CAS algorithms for parameter scans. However, its development lead to a modular structure allowing the implementation of further PS algorithms existing in the literature such as MLScan.

We demonstrated the utility of the library by performing multi-objective searches on test
functions and comparing the performance of different PS algorithms. In this connection,  b-CASTOR exhibited
superior sample efficiency while achieving a comprehensive characterisation of the satisfactory
parameter region. The experiments with the MLScan algorithm instead highlighted its robust exploration
capabilities, due to the stochastic nature of its policy. However, for BSM phenomenological studies,
a NN classifier needs to be implemented to enhance the sample efficiency of this  algorithm. (All ML-based such approaches were also demonstrated 
to be superior to the more standard MCMC-HS one.)
Additionally, we illustrated the application of the library in a real BSM  case study, fitting
the masses of the lightest Higgs particles in the $(B-L)$SSM to explain new physics signals
in $\gamma\gamma$, $\tau^+\tau^-$, and/or $b\bar b$ final states \cite{madiaz2024}.

Looking ahead, future work will focus on improving the \hepaid\ design and modularity to encourage
the development of new PS algorithms, possibly starting from combinations of existing
components: e.g., using the rejection sampling component from MLScan to sample from the
ECI policy function, which is used by CAS and b-CASTOR, could enhance 
exploration capabilities. This is, in fact, a direction currently under investigation. Another promising development  is
integrating \hepaid\ with other existing PS libraries, particularly those that
interface with a broad range of HEP software. Ultimately, the primary goal of \hepaid\ is to serve as a
versatile testing ground for developing and benchmarking PS
algorithms, with a focus on ML-driven approaches.

\section*{Acknowledgments}
\noindent
The work of SM is supported in part through the NExT Institute and the STFC Consolidated Grant No. ST/X000583/1. MAD and his work were supported by ANID BECAS DE DOCTORADO EN EL EXTRANJERO, BECAS CHILE 2020, 72210042.

\appendix

\bibliographystyle{elsarticle-num} 
\bibliography{main.bib}

\begin{thebibliography}{10}
\expandafter\ifx\csname url\endcsname\relax
  \def\url#1{\texttt{#1}}\fi
\expandafter\ifx\csname urlprefix\endcsname\relax\def\urlprefix{URL }\fi
\expandafter\ifx\csname href\endcsname\relax
  \def\href#1#2{#2} \def\path#1{#1}\fi

\bibitem{Brein_2005}
O.~Brein, \href{http://dx.doi.org/10.1016/j.cpc.2005.03.104}{Adaptive scanning—a proposal how to scan theoretical predictions over a multi-dimensional parameter space efficiently}, Computer Physics Communications 170~(1) (2005) 42–48.
\newblock \href {https://doi.org/10.1016/j.cpc.2005.03.104} {\path{doi:10.1016/j.cpc.2005.03.104}}.
\newline\urlprefix\url{http://dx.doi.org/10.1016/j.cpc.2005.03.104}

\bibitem{Hogg_2018}
D.~W. Hogg, D.~Foreman-Mackey, \href{http://dx.doi.org/10.3847/1538-4365/aab76e}{Data analysis recipes: Using markov chain monte carlo*}, The Astrophysical Journal Supplement Series 236~(1) (2018) 11.
\newblock \href {https://doi.org/10.3847/1538-4365/aab76e} {\path{doi:10.3847/1538-4365/aab76e}}.
\newline\urlprefix\url{http://dx.doi.org/10.3847/1538-4365/aab76e}

\bibitem{albert2024comparison}
J.~Albert, C.~Balazs, A.~Fowlie, W.~Handley, N.~Hunt-Smith, R.~R. de~Austri, M.~White, A comparison of bayesian sampling algorithms for high-dimensional particle physics and cosmology applications (2024).
\newblock \href {http://arxiv.org/abs/2409.18464} {\path{arXiv:2409.18464}}.

\bibitem{Speagle:2019ffr}
J.~S. Speagle, {A Conceptual Introduction to Markov Chain Monte Carlo Methods} (9 2019).
\newblock \href {http://arxiv.org/abs/1909.12313} {\path{arXiv:1909.12313}}.

\bibitem{feickert2021living}
M.~Feickert, B.~Nachman, A living review of machine learning for particle physics (2021).
\newblock \href {http://arxiv.org/abs/2102.02770} {\path{arXiv:2102.02770}}.

\bibitem{Baruah_2024}
R.~Baruah, S.~Mondal, S.~K. Patra, S.~Roy, \href{http://dx.doi.org/10.1140/epjs/s11734-024-01236-w}{Probing intractable beyond-standard-model parameter spaces armed with machine learning}, The European Physical Journal Special Topics (Jul. 2024).
\newblock \href {https://doi.org/10.1140/epjs/s11734-024-01236-w} {\path{doi:10.1140/epjs/s11734-024-01236-w}}.
\newline\urlprefix\url{http://dx.doi.org/10.1140/epjs/s11734-024-01236-w}

\bibitem{ren2019exploring}
J.~Ren, L.~Wu, J.~M. Yang, J.~Zhao, \href{https://www.sciencedirect.com/science/article/pii/S0550321319300938}{Exploring supersymmetry with machine learning}, Nuclear Physics B 943 (2019) 114613.
\newblock \href {https://doi.org/https://doi.org/10.1016/j.nuclphysb.2019.114613} {\path{doi:https://doi.org/10.1016/j.nuclphysb.2019.114613}}.
\newline\urlprefix\url{https://www.sciencedirect.com/science/article/pii/S0550321319300938}

\bibitem{Hammad_2023}
A.~Hammad, M.~Park, R.~Ramos, P.~Saha, \href{http://dx.doi.org/10.1016/j.cpc.2023.108902}{Exploration of parameter spaces assisted by machine learning}, Computer Physics Communications 293 (2023) 108902.
\newblock \href {https://doi.org/10.1016/j.cpc.2023.108902} {\path{doi:10.1016/j.cpc.2023.108902}}.
\newline\urlprefix\url{http://dx.doi.org/10.1016/j.cpc.2023.108902}

\bibitem{binjonaid2024multilabel}
M.~Binjonaid, Multi-label classification of parameter constraints in bsm extensions using deep learning (2024).
\newblock \href {http://arxiv.org/abs/2409.05453} {\path{arXiv:2409.05453}}.

\bibitem{Hollingsworth_2021}
J.~Hollingsworth, M.~Ratz, P.~Tanedo, D.~Whiteson, \href{http://dx.doi.org/10.1140/epjc/s10052-021-09941-9}{Efficient sampling of constrained high-dimensional theoretical spaces with machine learning}, The European Physical Journal C 81~(12) (Dec. 2021).
\newblock \href {https://doi.org/10.1140/epjc/s10052-021-09941-9} {\path{doi:10.1140/epjc/s10052-021-09941-9}}.
\newline\urlprefix\url{http://dx.doi.org/10.1140/epjc/s10052-021-09941-9}

\bibitem{Goodsell_2023}
M.~D. Goodsell, A.~Joury, \href{http://dx.doi.org/10.1140/epjc/s10052-023-11368-3}{Active learning bsm parameter spaces}, The European Physical Journal C 83~(4) (Apr. 2023).
\newblock \href {https://doi.org/10.1140/epjc/s10052-023-11368-3} {\path{doi:10.1140/epjc/s10052-023-11368-3}}.
\newline\urlprefix\url{http://dx.doi.org/10.1140/epjc/s10052-023-11368-3}

\bibitem{Goodsell2023BSMArtSA}
M.~D. Goodsell, A.~Joury, \href{https://www.sciencedirect.com/science/article/pii/S0010465523004022}{Bsmart: Simple and fast parameter space scans}, Vol. 297, 2024, p. 109057.
\newblock \href {https://doi.org/https://doi.org/10.1016/j.cpc.2023.109057} {\path{doi:https://doi.org/10.1016/j.cpc.2023.109057}}.
\newline\urlprefix\url{https://www.sciencedirect.com/science/article/pii/S0010465523004022}

\bibitem{PhysRevD.107.035004}
F.~Abreu~de Souza, M.~Crispim Rom\~ao, N.~F. Castro, M.~Nikjoo, W.~Porod, \href{https://link.aps.org/doi/10.1103/PhysRevD.107.035004}{Exploring parameter spaces with artificial intelligence and machine learning black-box optimization algorithms}, Phys. Rev. D 107 (2023) 035004.
\newblock \href {https://doi.org/10.1103/PhysRevD.107.035004} {\path{doi:10.1103/PhysRevD.107.035004}}.
\newline\urlprefix\url{https://link.aps.org/doi/10.1103/PhysRevD.107.035004}

\bibitem{romao2024combining}
J.~C. Romão, M.~C. Romão, Combining evolutionary strategies and novelty detection to go beyond the alignment limit of the $z_3$ 3hdm (2024).
\newblock \href {http://arxiv.org/abs/2402.07661} {\path{arXiv:2402.07661}}.

\bibitem{balazs_comparison_2021}
C.~Balázs, M.~van Beekveld, S.~Caron, B.~M. Dillon, B.~Farmer, A.~Fowlie, E.~C. Garrido-Merchán, W.~Handley, L.~Hendriks, G.~Jóhannesson, A.~Leinweber, J.~Mamužić, G.~D. Martinez, S.~Otten, R.~R. de~Austri, P.~Scott, Z.~Searle, B.~Stienen, J.~Vanschoren, M.~White, {The DarkMachines High Dimensional Sampling Group}, \href{https://doi.org/10.1007/JHEP05(2021)108}{A comparison of optimisation algorithms for high-dimensional particle and astrophysics applications}, Journal of High Energy Physics 2021~(5) (2021) 108.
\newblock \href {https://doi.org/10.1007/JHEP05(2021)108} {\path{doi:10.1007/JHEP05(2021)108}}.
\newline\urlprefix\url{https://doi.org/10.1007/JHEP05(2021)108}

\bibitem{madiaz2024}
M.~A. Diaz, G.~Cerro, S.~Dasmahapatra, S.~Moretti, \href{https://arxiv.org/abs/2404.18653}{Bayesian active search on parameter space: a 95 gev spin-0 resonance in the ($b-l$)ssm} (2024).
\newblock \href {http://arxiv.org/abs/2404.18653} {\path{arXiv:2404.18653}}.
\newline\urlprefix\url{https://arxiv.org/abs/2404.18653}

\bibitem{Staub:2011dp}
F.~Staub, T.~Ohl, W.~Porod, C.~Speckner, {A Tool Box for Implementing Supersymmetric Models}, Comput. Phys. Commun. 183 (2012) 2165--2206.
\newblock \href {http://arxiv.org/abs/1109.5147} {\path{arXiv:1109.5147}}, \href {https://doi.org/10.1016/j.cpc.2012.04.013} {\path{doi:10.1016/j.cpc.2012.04.013}}.

\bibitem{Staub:2019xhl}
F.~Staub, {xBIT: an easy to use scanning tool with machine learning abilities} (6 2019).
\newblock \href {http://arxiv.org/abs/1906.03277} {\path{arXiv:1906.03277}}.

\bibitem{Shang:2023gfy}
L.~Shang, Y.~Zhang, {EasyScan\_HEP: A tool for connecting programs to scan the parameter space of physics models}, Comput. Phys. Commun. 296 (2024) 109027.
\newblock \href {http://arxiv.org/abs/2304.03636} {\path{arXiv:2304.03636}}, \href {https://doi.org/10.1016/j.cpc.2023.109027} {\path{doi:10.1016/j.cpc.2023.109027}}.

\bibitem{Goodsell:2023iac}
M.~D. Goodsell, A.~Joury, {BSMArt: Simple and fast parameter space scans}, Comput. Phys. Commun. 297 (2024) 109057.
\newblock \href {http://arxiv.org/abs/2301.01154} {\path{arXiv:2301.01154}}, \href {https://doi.org/10.1016/j.cpc.2023.109057} {\path{doi:10.1016/j.cpc.2023.109057}}.

\bibitem{Staub:2008uz}
F.~Staub, {SARAH} (6 2008).
\newblock \href {http://arxiv.org/abs/0806.0538} {\path{arXiv:0806.0538}}.

\bibitem{Staub_2014}
F.~Staub, \href{http://dx.doi.org/10.1016/j.cpc.2014.02.018}{Sarah 4: A tool for (not only susy) model builders}, Computer Physics Communications 185~(6) (2014) 1773–1790.
\newblock \href {https://doi.org/10.1016/j.cpc.2014.02.018} {\path{doi:10.1016/j.cpc.2014.02.018}}.
\newline\urlprefix\url{http://dx.doi.org/10.1016/j.cpc.2014.02.018}

\bibitem{garnett2012bayesian}
R.~Garnett, Y.~Krishnamurthy, X.~Xiong, J.~Schneider, R.~Mann, Bayesian optimal active search and surveying, in: Proceedings of the 29th International Conference on International Conference on Machine Learning, ICML'12, Omnipress, Madison, WI, USA, 2012, p. 843–850.

\bibitem{reback2020pandas}
T.~pandas~development team, \href{https://doi.org/10.5281/zenodo.3509134}{pandas-dev/pandas: Pandas} (Feb. 2020).
\newblock \href {https://doi.org/10.5281/zenodo.3509134} {\path{doi:10.5281/zenodo.3509134}}.
\newline\urlprefix\url{https://doi.org/10.5281/zenodo.3509134}

\bibitem{paszke2019pytorch}
A.~Paszke, S.~Gross, F.~Massa, A.~Lerer, J.~Bradbury, G.~Chanan, T.~Killeen, Z.~Lin, N.~Gimelshein, L.~Antiga, A.~Desmaison, A.~Köpf, E.~Yang, Z.~DeVito, M.~Raison, A.~Tejani, S.~Chilamkurthy, B.~Steiner, L.~Fang, J.~Bai, S.~Chintala, Pytorch: An imperative style, high-performance deep learning library (2019).
\newblock \href {http://arxiv.org/abs/1912.01703} {\path{arXiv:1912.01703}}.

\bibitem{Abdelalim:2020xfk}
A.~A. Abdelalim, B.~Das, S.~Khalil, S.~Moretti, {Di-photon decay of a light Higgs state in the BLSSM}, Nucl. Phys. B 985 (2022) 116013.
\newblock \href {http://arxiv.org/abs/2012.04952} {\path{arXiv:2012.04952}}, \href {https://doi.org/10.1016/j.nuclphysb.2022.116013} {\path{doi:10.1016/j.nuclphysb.2022.116013}}.

\bibitem{CMS:2023yay}
{Search for a standard model-like Higgs boson in the mass range between 70 and 110$~\mathrm{GeV}$ in the diphoton final state in proton-proton collisions at $\sqrt{s}=13~\mathrm{TeV}$} (2023).

\bibitem{Porod2003SPhenoAP}
W.~Porod, \href{https://api.semanticscholar.org/CorpusID:7905927}{Spheno, a program for calculating supersymmetric spectra, susy particle decays and susy particle production at e+e- colliders}, Computer Physics Communications 153 (2003) 275--315.
\newline\urlprefix\url{https://api.semanticscholar.org/CorpusID:7905927}

\bibitem{Porod_2012}
W.~Porod, F.~Staub, \href{http://dx.doi.org/10.1016/j.cpc.2012.05.021}{Spheno 3.1: extensions including flavour, cp-phases and models beyond the mssm}, Computer Physics Communications 183~(11) (2012) 2458--2469.
\newblock \href {https://doi.org/10.1016/j.cpc.2012.05.021} {\path{doi:10.1016/j.cpc.2012.05.021}}.
\newline\urlprefix\url{http://dx.doi.org/10.1016/j.cpc.2012.05.021}

\bibitem{Bechtle_2010}
P.~Bechtle, O.~Brein, S.~Heinemeyer, G.~Weiglein, K.~Williams, \href{http://dx.doi.org/10.1016/j.cpc.2009.09.003}{Higgsbounds: Confronting arbitrary higgs sectors with exclusion bounds from lep and the tevatron}, Computer Physics Communications 181~(1) (2010) 138–167.
\newblock \href {https://doi.org/10.1016/j.cpc.2009.09.003} {\path{doi:10.1016/j.cpc.2009.09.003}}.
\newline\urlprefix\url{http://dx.doi.org/10.1016/j.cpc.2009.09.003}

\bibitem{Bechtle_2014}
P.~Bechtle, S.~Heinemeyer, O.~Stål, T.~Stefaniak, G.~Weiglein, \href{http://dx.doi.org/10.1140/epjc/s10052-013-2711-4}{Higgssignals: Confronting arbitrary higgs sectors with measurements at the tevatron and the lhc}, The European Physical Journal C 74~(2) (Feb. 2014).
\newblock \href {https://doi.org/10.1140/epjc/s10052-013-2711-4} {\path{doi:10.1140/epjc/s10052-013-2711-4}}.
\newline\urlprefix\url{http://dx.doi.org/10.1140/epjc/s10052-013-2711-4}

\bibitem{Alwall:2014hca}
J.~Alwall, R.~Frederix, S.~Frixione, V.~Hirschi, F.~Maltoni, O.~Mattelaer, H.~S. Shao, T.~Stelzer, P.~Torrielli, M.~Zaro, {The automated computation of tree-level and next-to-leading order differential cross sections, and their matching to parton shower simulations}, JHEP 07 (2014) 079.
\newblock \href {http://arxiv.org/abs/1405.0301} {\path{arXiv:1405.0301}}, \href {https://doi.org/10.1007/JHEP07(2014)079} {\path{doi:10.1007/JHEP07(2014)079}}.

\bibitem{malkomes2021beyond}
G.~Malkomes, B.~Cheng, E.~H. Lee, M.~Mccourt, \href{https://proceedings.mlr.press/v139/malkomes21a.html}{Beyond the pareto efficient frontier: Constraint active search for multiobjective experimental design}, in: M.~Meila, T.~Zhang (Eds.), Proceedings of the 38th International Conference on Machine Learning, Vol. 139 of Proceedings of Machine Learning Research, PMLR, 2021, pp. 7423--7434.
\newline\urlprefix\url{https://proceedings.mlr.press/v139/malkomes21a.html}

\bibitem{Metropolis:1953am}
N.~Metropolis, A.~W. Rosenbluth, M.~N. Rosenbluth, A.~H. Teller, E.~Teller, {Equation of state calculations by fast computing machines}, J. Chem. Phys. 21 (1953) 1087--1092.
\newblock \href {https://doi.org/10.1063/1.1699114} {\path{doi:10.1063/1.1699114}}.

\bibitem{Hastings:1970aa}
W.~K. Hastings, {Monte Carlo Sampling Methods Using Markov Chains and Their Applications}, Biometrika 57 (1970) 97--109.
\newblock \href {https://doi.org/10.1093/biomet/57.1.97} {\path{doi:10.1093/biomet/57.1.97}}.

\bibitem{Allanach:2008qq}
B.~C. Allanach, C.~Balazs, G.~Belanger, M.~Bernhardt, F.~Boudjema, D.~Choudhury, K.~Desch, U.~Ellwanger, P.~Gambino, R.~Godbole, T.~Goto, J.~Guasch, M.~Guchait, T.~Hahn, S.~Heinemeyer, C.~Hugonie, T.~Hurth, S.~Kraml, S.~Kreiss, J.~Lykken, F.~Moortgat, S.~Moretti, S.~Penaranda, T.~Plehn, W.~Porod, A.~Pukhov, P.~Richardson, M.~Schumacher, L.~Silvestrini, P.~Skands, P.~Slavich, M.~Spira, G.~Weiglein, P.~Wienemann, {SUSY Les Houches Accord 2}, Comput. Phys. Commun. 180 (2009) 8--25.
\newblock \href {http://arxiv.org/abs/0801.0045} {\path{arXiv:0801.0045}}, \href {https://doi.org/10.1016/j.cpc.2008.08.004} {\path{doi:10.1016/j.cpc.2008.08.004}}.

\bibitem{Darme:2023hni}
L.~Darm{\'e}, C.~Degrande, C.~Duhr, B.~Fuks, M.~Goodsell, G.~Heinrich, V.~Hirschi, S.~H{\"o}che, M.~H{\"o}fer, J.~Isaacson, O.~Mattelaer, T.~Ohl, D.~Pagani, J.~Reuter, P.~Richardson, S.~Schumann, H.-S. Shao, F.~Siegert, M.~Zaro, {UFO 2.0 -- The Universal Feynman Output format}, Eur. Phys. J. C 83~(7) (2023) 631.
\newblock \href {http://arxiv.org/abs/2304.09883} {\path{arXiv:2304.09883}}, \href {https://doi.org/10.1140/epjc/s10052-023-11780-9} {\path{doi:10.1140/epjc/s10052-023-11780-9}}.

\bibitem{CMS-PAS-HIG-21-001}
\href{https://cds.cern.ch/record/2803739}{{Searches for additional Higgs bosons and vector leptoquarks in $\tau\tau$ final states in proton-proton collisions at $\sqrt{s}=13~\mathrm{TeV}$}}, Tech. rep., CERN, Geneva (2022).
\newline\urlprefix\url{https://cds.cern.ch/record/2803739}

\bibitem{LEPWorkingGroupforHiggsbosonsearches:2003ing}
R.~Barate, et~al., {Search for the standard model Higgs boson at LEP}, Phys. Lett. B 565 (2003) 61--75.
\newblock \href {http://arxiv.org/abs/hep-ex/0306033} {\path{arXiv:hep-ex/0306033}}, \href {https://doi.org/10.1016/S0370-2693(03)00614-2} {\path{doi:10.1016/S0370-2693(03)00614-2}}.

\bibitem{atlas2012}
G.~Aad, et~al., \href{https://www.sciencedirect.com/science/article/pii/S037026931200857X}{Observation of a new particle in the search for the standard model higgs boson with the atlas detector at the lhc}, Physics Letters B 716~(1) (2012) 1--29.
\newblock \href {https://doi.org/https://doi.org/10.1016/j.physletb.2012.08.020} {\path{doi:https://doi.org/10.1016/j.physletb.2012.08.020}}.
\newline\urlprefix\url{https://www.sciencedirect.com/science/article/pii/S037026931200857X}

\bibitem{cms2012}
S.~Chatrchyan, et~al., \href{https://www.sciencedirect.com/science/article/pii/S0370269312008581}{Observation of a new boson at a mass of 125 gev with the cms experiment at the lhc}, Physics Letters B 716~(1) (2012) 30--61.
\newblock \href {https://doi.org/https://doi.org/10.1016/j.physletb.2012.08.021} {\path{doi:https://doi.org/10.1016/j.physletb.2012.08.021}}.
\newline\urlprefix\url{https://www.sciencedirect.com/science/article/pii/S0370269312008581}

\bibitem{Bahl:2022igd}
H.~Bahl, T.~Biek\"otter, S.~Heinemeyer, C.~Li, S.~Paasch, G.~Weiglein, J.~Wittbrodt, {HiggsTools: BSM scalar phenomenology with new versions of HiggsBounds and HiggsSignals}, Comput. Phys. Commun. 291 (2023) 108803.
\newblock \href {http://arxiv.org/abs/2210.09332} {\path{arXiv:2210.09332}}, \href {https://doi.org/10.1016/j.cpc.2023.108803} {\path{doi:10.1016/j.cpc.2023.108803}}.

\bibitem{Skands:2003cj}
P.~Z. Skands, et~al., {SUSY Les Houches accord: Interfacing SUSY spectrum calculators, decay packages, and event generators}, JHEP 07 (2004) 036.
\newblock \href {http://arxiv.org/abs/hep-ph/0311123} {\path{arXiv:hep-ph/0311123}}, \href {https://doi.org/10.1088/1126-6708/2004/07/036} {\path{doi:10.1088/1126-6708/2004/07/036}}.

\bibitem{buckley2015pyslhapythonicinterfacesusy}
A.~Buckley, \href{https://arxiv.org/abs/1305.4194}{Pyslha: a pythonic interface to susy les houches accord data} (2015).
\newblock \href {http://arxiv.org/abs/1305.4194} {\path{arXiv:1305.4194}}.
\newline\urlprefix\url{https://arxiv.org/abs/1305.4194}

\bibitem{Staub_2019_xslha}
F.~Staub, \href{http://dx.doi.org/10.1016/j.cpc.2019.03.013}{xslha: An les houches accord reader for python and mathematica}, Computer Physics Communications 241 (2019) 132–138.
\newblock \href {https://doi.org/10.1016/j.cpc.2019.03.013} {\path{doi:10.1016/j.cpc.2019.03.013}}.
\newline\urlprefix\url{http://dx.doi.org/10.1016/j.cpc.2019.03.013}

\bibitem{head_2021_5565057}
T.~Head, M.~Kumar, H.~Nahrstaedt, G.~Louppe, I.~Shcherbatyi, \href{https://doi.org/10.5281/zenodo.5565057}{scikit-optimize/scikit-optimize} (Oct. 2021).
\newblock \href {https://doi.org/10.5281/zenodo.5565057} {\path{doi:10.5281/zenodo.5565057}}.
\newline\urlprefix\url{https://doi.org/10.5281/zenodo.5565057}

\bibitem{omegaconf2019}
O.~Yadan, J.~Sommer-Simpson, O.~Delalleau, \href{https://github.com/omry/omegaconf}{omegaconf} (2019).
\newline\urlprefix\url{https://github.com/omry/omegaconf}

\bibitem{balandat2020botorch}
M.~Balandat, B.~Karrer, D.~R. Jiang, S.~Daulton, B.~Letham, A.~G. Wilson, E.~Bakshy, \href{http://arxiv.org/abs/1910.06403}{{BoTorch: A Framework for Efficient Monte-Carlo Bayesian Optimization}}, in: Advances in Neural Information Processing Systems 33, 2020.
\newline\urlprefix\url{http://arxiv.org/abs/1910.06403}

\bibitem{NEURIPS2018_27e8e171}
J.~Gardner, G.~Pleiss, K.~Q. Weinberger, D.~Bindel, A.~G. Wilson, Gpytorch: Blackbox matrix-matrix gaussian process inference with gpu acceleration, in: S.~Bengio, H.~Wallach, H.~Larochelle, K.~Grauman, N.~Cesa-Bianchi, R.~Garnett (Eds.), Advances in Neural Information Processing Systems, Vol.~31, Curran Associates, Inc., 2018.

\bibitem{optuna_2019}
T.~Akiba, S.~Sano, T.~Yanase, T.~Ohta, M.~Koyama, Optuna: A next-generation hyperparameter optimization framework, in: Proceedings of the 25th {ACM} {SIGKDD} International Conference on Knowledge Discovery and Data Mining, 2019.

\end{thebibliography}





\end{document}